\documentclass[prb,prb,aps,superscriptaddress,twocolumn,showpacs,preprintnumbers,amsmath,amssymb]{revtex4-1}
\usepackage{epsf}
\usepackage{color}

\begin{document}
\draft
\newcommand{\ve}[1]{\boldsymbol{#1}}

\title{Oxygen vacancies at titanate interfaces: \\ 
two-dimensional magnetism and orbital reconstruction}

\author{N.~Pavlenko}
\altaffiliation{On leave from the Institute for Condensed Matter Physics, NAS, 79011 Lviv, Ukraine} 
\affiliation{Center for Electronic Correlations and Magnetism, Theoretical Physics III, Institute of Physics, University of Augsburg, 86135 Augsburg, Germany}
\affiliation{Center for Electronic Correlations and Magnetism, Experimental Physics VI, Institute of Physics, University of Augsburg, 86135 Augsburg, Germany}
\author{T.~Kopp$^2$}
\noaffiliation
\author{E.Y.~Tsymbal}
\affiliation{Department of Physics and Astronomy, Nebraska Center for Materials and Nanoscience, University of Nebraska, Lincoln, Nebraska 68588-0299, USA}
\author{J.~Mannhart}
\affiliation{Max Planck Institute for Solid State Research, 70569 Stuttgart, Germany}
\author{G.A.~Sawatzky}
\affiliation{Department of Physics and Astronomy, University of British Columbia, Vancouver, Canada V6T1Z1}

\begin{abstract}
We show that oxygen vacancies at titanate interfaces induce a complex
multiorbital reconstruction which involves a lowering of the local symmetry 
and an inversion of $t_{2g}$ and $e_{g}$ orbitals resulting in the  occupation 
of the  $e_{g}$ orbitals of Ti atoms neighboring the O vacancy.
The orbital reconstruction 
depends strongly on the clustering of O vacancies and can be accompanied by a magnetic splitting between the local $e_g$ orbitals with lobes directed towards the vacancy and interface $d_{xy}$ orbitals. The reconstruction generates a two-dimensional interface magnetic state not observed in bulk SrTiO$_3$. Using generalized gradient 
approximation (LSDA) with intra-atomic
Coulomb repulsion (GGA+U), we find that this magnetic state is common for
titanate surfaces and interfaces.
\end{abstract}


\maketitle

\section{Introduction}

Charged impurities can change the electronic properties of insulating materials qualitatively.
For example, doping of SrTiO$_3$ with Nb leads to semiconducting properties and eventually to a
superconducting transition below $0.3$~K \cite{schooley}. The recent discovery of the metallic,
superconducting and magnetic states at the interface between the bulk insulators SrTiO$_3$(STO) and LaAlO$_3$(LAO) triggered 
an intense exploration of electronic reconstruction and the 
role of impurities in the formation of a conducting and sometimes magnetic state at these interfaces \cite{ohtomo,brinkman,kalabukhov,dikin,li,ariando,salluzzo,tsymbal,pavlenko,bert,michaeli,muller,
pentcheva,pavlenko4,zhong,bristowe,sing}.

Two distinct doping mechanisms, controlled by the growth conditions of the samples, appear to be responsible for the formation of interface electron carriers
\cite{eckstein,thiel,barthelemy,siemons,herranz}. 
The intrinsic polar charge doping in the heterostructures
prepared at high oxygen pressures ($10^{-4}$~mbar) suggests doping with an upper carrier density limit of $3\cdot 10^{14}$~cm$^{-2}$
which corresponds to 0.5 electrons per interface unit cell (uc) and compensates the 
interface polar discontinuity. The resulting two-dimensional electron liquid forms 
in a nm-thick interface layer \cite{thiel,barthelemy,janicka,eom,demkov,chen,bark} and can be a possible source for superconducting properties tuned by gate electrostatic fields \cite{thiel,reyren}.       
The extrinsic mechanism of charge doping due to oxygen vacancies 
dominates in the LAO/STO samples grown at lower oxygen pressures ($10^{-6}$~mbar)
reported in Refs.~\onlinecite{ohtomo,kalabukhov}. Remarkably, the oxygen reduced heterostructures are characterized by higher charge densities of
$10^{16}- 10^{17}$~cm$^{-2}$ with the charge spatially extended into the $\mu$m-thick interfacial 
layer inside the STO \cite{barthelemy}. 
Recent cathode and photoluminescence experiments \cite{kalabukhov} provide direct 
evidence for oxygen vacancies in the STO-substrate.
The oxygen vacancies can diffuse in STO with activation energy of $0.75$~eV, 
which is comparable to the activation energy $1$~eV of the vacancy migration through the LAO layers\cite{chen2}. 
At each oxygen vacancy in the TiO$_2$ layer, two electrons are weakly bound 
to retain charge neutrality of a configuration in which the two titanium ions in the dimer share two electrons.
Due to these moderately bound electrons the oxygen vacancies act as n-type dopants.
We note though that in STO, these O-vacancy induced, weakly bound electrons do not contribute to resolve
the polar discontinuity problem since they simply are there to produce charge neutrality upon removing the oxygens. The O-vacancy induced electrons are bound to the effective $2+$ charge of the vacancy. Below we argue that these electrons enhance the interface conductivity considerably.

A critical issue related to the electron concentration is the electric conductivity of LAO/STO heterointerfaces.
The experimental measurements show that the conductivity of n-type LAO/STO bilayers increases with lowering
the oxygen partial pressure $p_{\rm O_2}$ and exhibits an abrupt jump from $10^{-2}$ to $10^{2}$~$\Omega^{-1}/ 
\square$ when p$_{\rm O_2}$ changes from $10^{-5}$ to $10^{-6}$~mbar~\cite{kalabukhov,ohtomo}. 
The high conductivity shown by samples grown at low oxygen pressures is caused by a large amount of oxygen vacancies which act as electron-donors in such samples \cite{ohtomo,muller}.

The local electronic state of the vacancy-containing transition metal oxides can exhibit 
extraordinary properties not observed in chemically stoichiometric materials. Recent
first principle studies demonstrate a vacancy-related magnetic
exchange splitting in nonmagnetic materials like CaO or LAO/STO~\cite{elfimov,pavlenko}. 
For LAO/STO, calculations of the vacancy formation energies\cite{zhang,pavlenko} suggest a
predominant positioning of O-vacancies in the top AlO$_2$ surfaces. The 
energy barrier for transport of O-vacancy
from the interface TiO$_2$ to the surface AlO$_2$ layer strongly depends on the vacancy concentration~\cite{yunli,pavlenko}.

Recent STM, cathode luminescence studies and conductivity measurements provide strong support for
clustering of the oxygen vacancies in STO \cite{muller,kalabukhov}. Moreover, the STM experiments
and DFT+$U$ studies suggest the possibility that oxygen vacancies form pieces of linear chains, which appear to be more energetically favourable \cite{cuong} as compared to the isolated vacancies\cite{luo,shanthi,ricci}. 

In this work, we explore the electronic structure of the first titanate surface/interface layers considering different configurations and concentrations of oxygen vacancies. 
In bulk SrTiO$_3$, the octahedral crystal ligand field, covalency and Coulomb repulsion result in higher energies of Ti $e_g$ states 
as compared to $t_{2g}$ states with a splitting of about $2.5$~eV\cite{wolfram,kahn,mattheiss}.  
Here we find that in contrast to the bulk stoichiometric SrTiO$_3$, each Ti-O$_v$-Ti dimer contains 
missing hybridization links which strongly lower the energy of the $d$ orbitals with lobes pointing to the vacancy. As a result, we obtain a splitting of the doubly degenerate $e_g$ orbitals and the triply degenerate $t_{2g}$ orbitals into orbital doublets and singlets.  
In the pure material, these $e_g$-like antibonding orbitals are strongly pushed up due to the strong hybrization with the lower-energy O $2p$ orbitals. 
In the oxygen-reduced LAO/STO, the missing local covalent bonding and the local symmetry lowering results in a new kind of orbital reconstruction at titanate interfaces: a shift and partial occupation of the vacancy-directed Ti $e_g$-orbitals accompanied by their magnetic 
splitting and mixing with the $3d_{xy}$ states. Our calculations of SrTiO$_3$-surfaces and LAO/STO interfaces demonstrate the universal character of the orbital reconstruction of the titanates due to surface/interface oxygen vacancies. In the linear clusters of vacancy stripes, in which each Ti is accompanied by two O-vacancies the $e_g$ orbital state pointing along the Ti vacancy direction, is further strongly lowered. 
In these $e_g$ states, both spin states are occupied, although a small amount of spin polarization is also present.
The high spin polarization occurs predominantly in the $t_{2g}$-states.

We also show that the electronic states of the oxygen-reduced STO and LAO layers 
are physically different. The vacancies
in TiO$_2$ layers release extra $3d$ electrons which are confined to the quasi-two-dimensional layer of the electron
liquid at the LAO/STO interface. 
In contrast, the oxygen vacancies in the LAO overlayer produce 
local electronic states of a mixed $sp$ character which are redistributed between the surface AlO$_2$  
and interface TiO$_2$ layers resulting in the confinement of electrons in the parallel, surface and 
interface 2D layers with different effective masses, a concept which has been recently discussed
in Refs.~\onlinecite{popovic,zhou}. We find that for small concentrations of oxygen vacancies $c_V\le 1/8$ at the AlO$_2$ surface, the vacancy released electrons are transferred to the LAO/STO interface and compensate the interface polar-discontinuity, which also leads to an insulating surface state.

\section{Oxygen vacancies at LAO/STO interfaces}

\begin{figure}[tbp]
\epsfxsize=8.5cm {\epsfclipon\epsffile{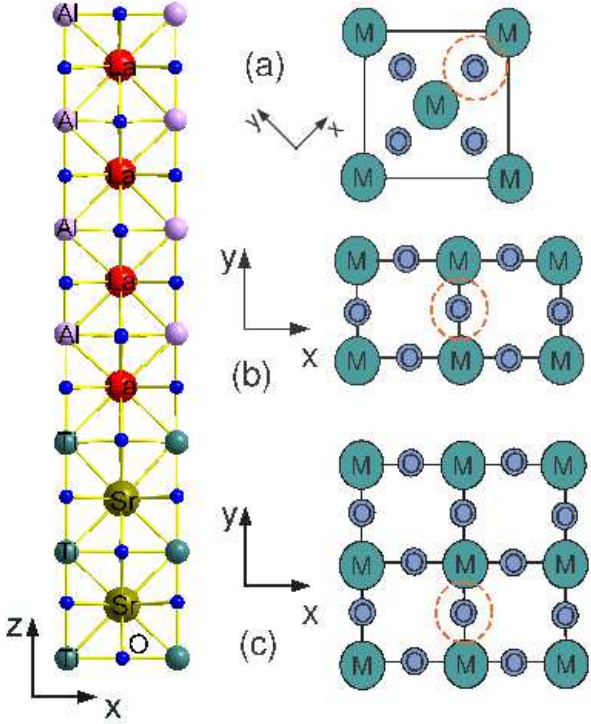}}
\caption{Schematic view of the SrTiO$_3$/LaAlO$_3$ heterostructure.
The supercell contains a 4 uc thick LaAlO$_3$ layer deposited on a 2.5 uc thick SrTiO$_3$ slab.
The full supercell consists of two symmetric parts of the depicted structure and a vacuum layer of 13~\AA.
The structures on the right side show M$_n$O$_{2n}$ (M=Ti, Al)-plaquettes (a) with one 
eliminated O(2a,a) atom (dimerized vacancy), (b) with eliminated O(0.5a,0.5a) (chained vacancy) and (c) with eliminated O(0.5a, 0.25a), generated for the study of the systems 
with O-vacancies. The position of an O-vacancy is identified by a
red dashed circle.
}
\label{fig1}
\end{figure}

We use supercells in density functional theory to explore the charge 
and spin state around oxygen vacancies at LAO/STO heterointerfaces.
The supercells contain a 4 uc thick LAO overlayer deposited on a STO layer
of a thickness between 1 uc and 6 uc. 
The LAO-STO-LAO structures are separated by a 13~\AA\, thick vacuum sheet.
In the supercell, oxygen vacancies can be located in the interface TiO$_2$ layer or in one
of the AlO$_2$-planes of LAO.
Cells with three types of oxygen vacancy configurations in MO$_2$ (M=Ti, Al) are sketched in
Fig.~\ref{fig1}(a), (b) and (c). Configuration (a) contains oxygen vacancies in Ti-O$_v$-Ti dimers, with each vacancy
located between two Ti atoms in a ($\sqrt{2} \times \sqrt{2}$) uc
so that each Ti in the TiO$_2$ has exactly one nearest vacancy.  We refer to this as a Ti dimer structure.
In contrast, the configuration (b) is introduced by removing the oxygen atom
O($a/2$, $b/2$) in the center of the ($2\times 1$) M$_2$O$_4$-plaquette, a configuration which results in 
stripes of vacancies in the $y$-direction. The doping level of O-vacancies 
in configurations (a) and (b)
corresponds to $c_V=1/4$ (25\%) of vacancies per each four oxygen atomic sites. 
The lowest density of O-vacancies (one vacancy per 8 oxygen sites) 
is represented by the configuration (c) in Fig.~\ref{fig1}, which as in configuration (a) has Ti-O$_v$-Ti dimers but the density of vacancies is a factor of two smaller. 

The density functional calculations were performed using the Generalized Gradient Approximation (GGA) in the
Perdew-Burke-Ernzerhof pseudopotential implementation \cite{pbe} in the Quantum Espresso (QE)
package \cite{qe}. We use a kinetic energy cutoff of 640~eV, and the Brillouin zone
of the 106- to 166-atom supercells is sampled with 5$\times$5$\times$1 to 9$\times$9$\times$1 $\ve{k}$-point
grids. An increase of the k-point mesh from (5x5x1) to (7x7x1) leads
to a negligibly small change of the total energy by 0.005~Ry and to an increase
of the Ti magnetic moments by small values of about 0.05~$\mu_B$ for the case that O-vacancies are present.
In our calculations we account for a local Coulomb repulsion of Ti $3d$ electrons
by employing a GGA+$U$ approach with $U_{\rm Ti}=2$~eV~\cite{breitschaft}.
The supercells have been structurally relaxed by a
combination of the  optimization procedures of the full potential WIEN2k-package and the pseudopotential QE
package \cite{wien2k,qe}. The in-plane lattice constants have been fixed to their room-temperature bulk-STO cubic values ($a_{{\rm STO}}=b_{{\rm STO}}=3.905~$\AA). Although low-temperature SrTiO$_3$ is tetragonal with the lattice parameters $a=3.896$\AA\, and $c=3.899$~\AA, in our studies we assume that the small difference between the lattice constants should not seriously affect the state of oxygen vacancies.

\begin{figure}[htbp]
\epsfxsize=7.0cm {\epsfclipon\epsffile{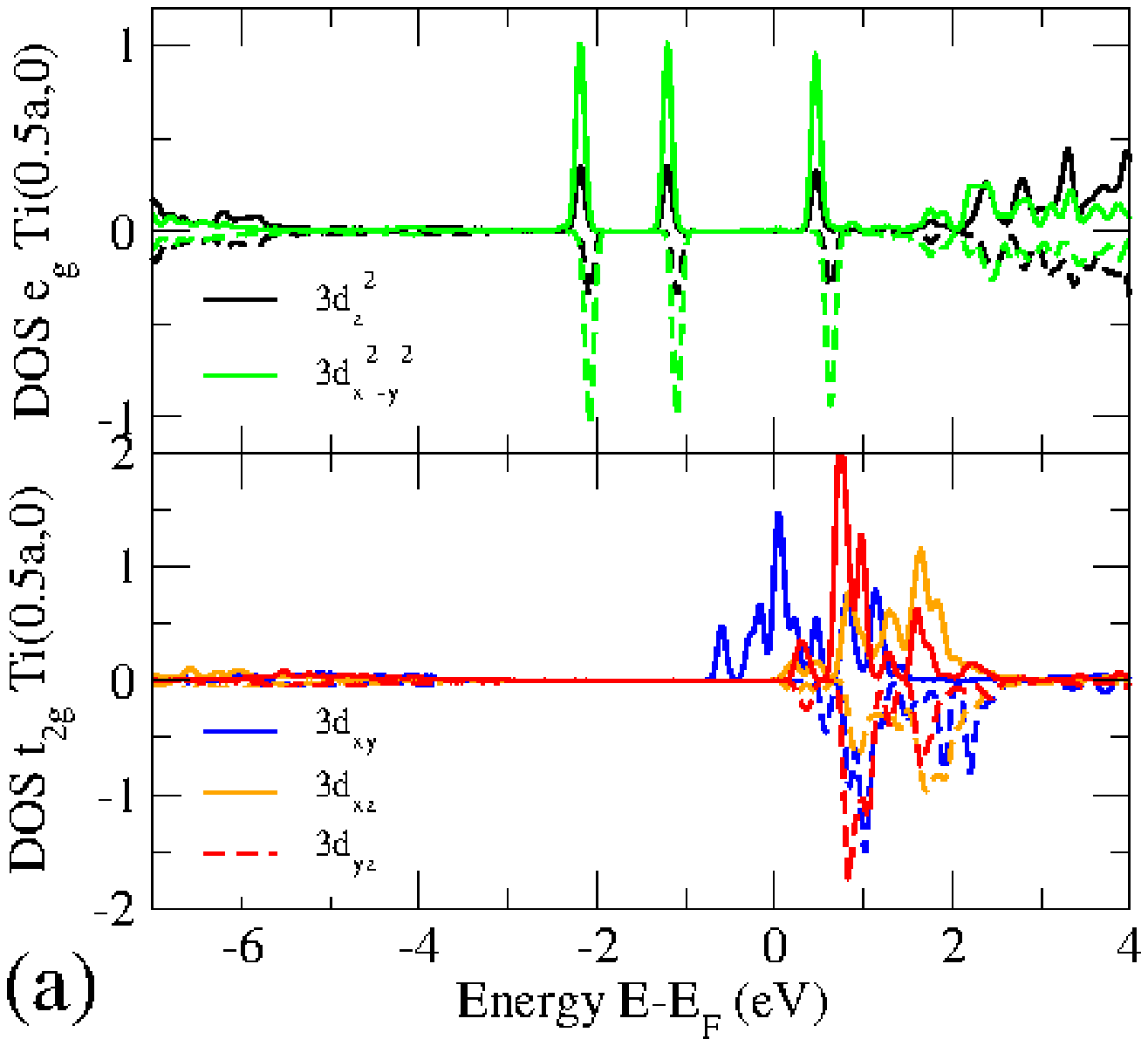}}
\epsfxsize=7.0cm {\epsfclipon\epsffile{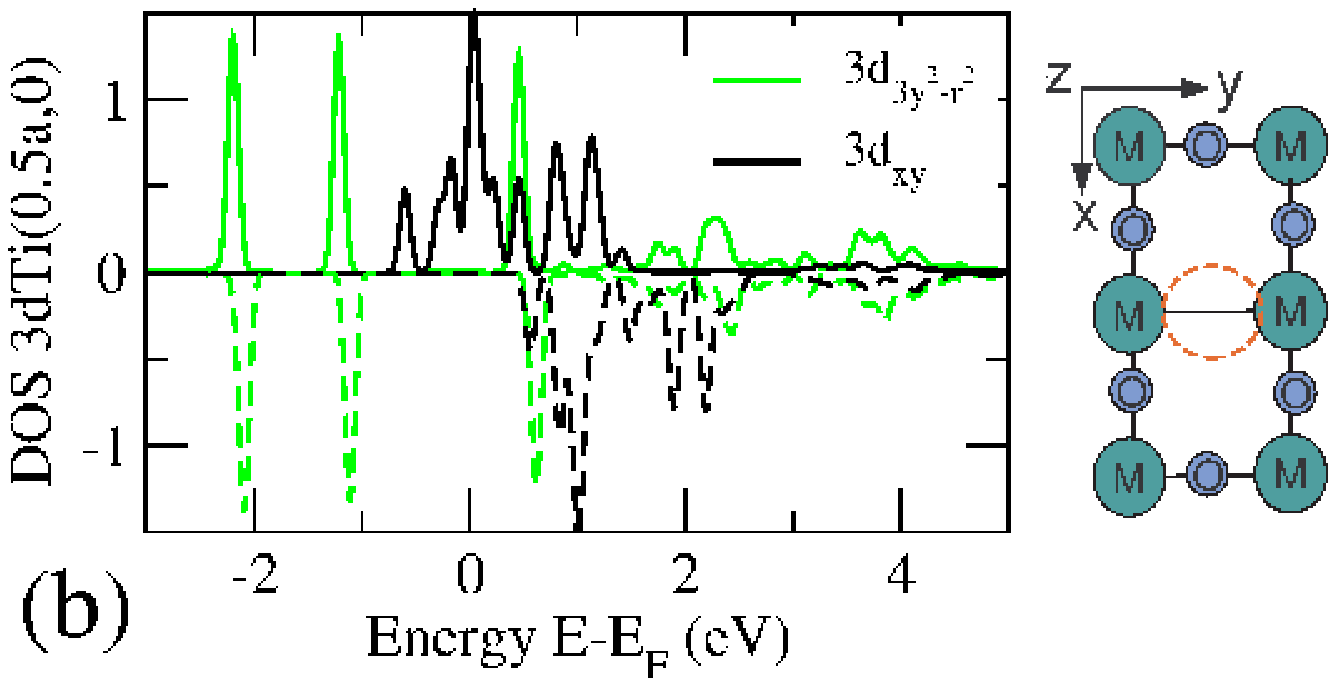}}
\epsfxsize=7.0cm {\epsfclipon\epsffile{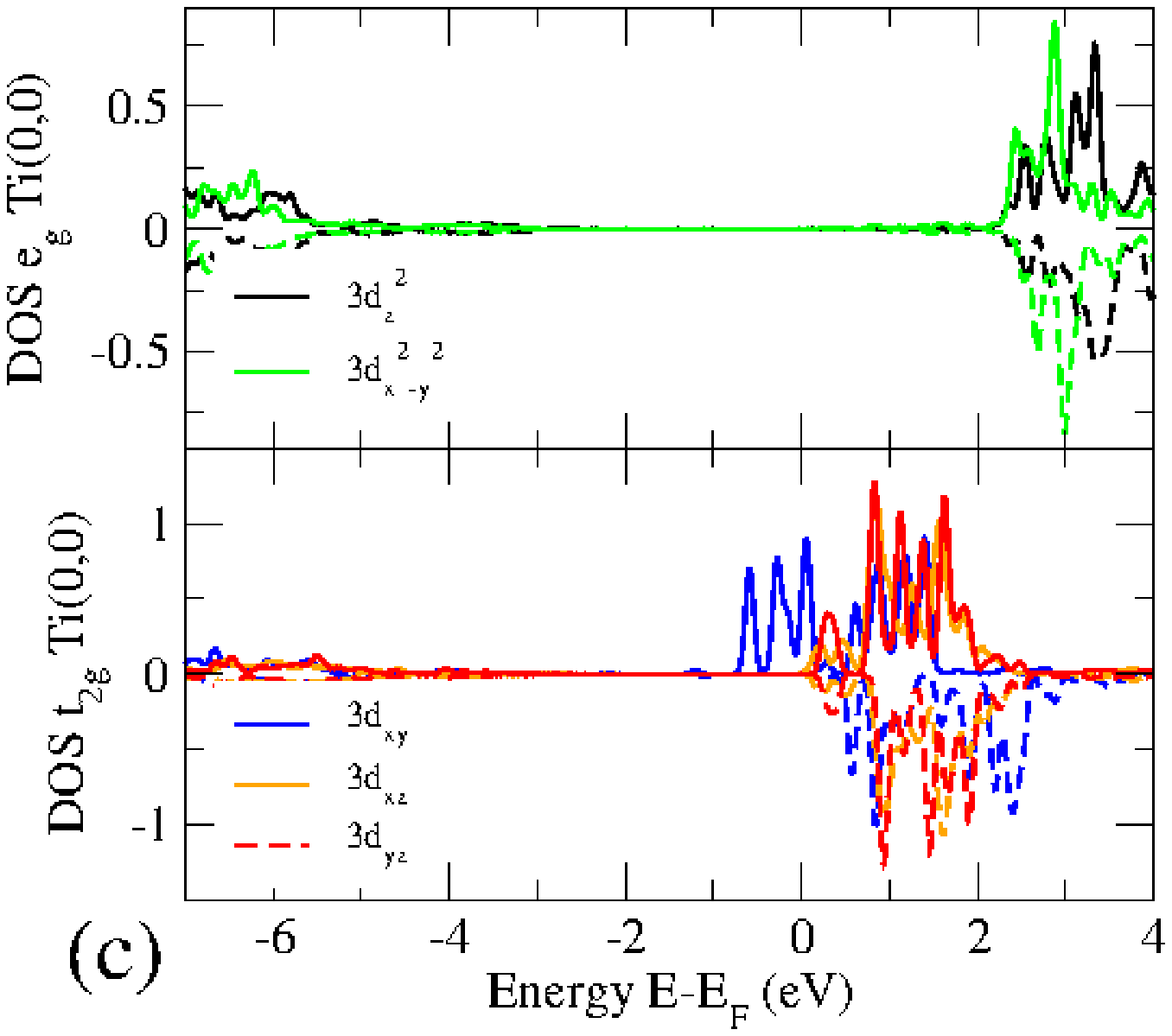}}
\caption{Spin-polarized orbital-projected densities of states for the chained vacancy configuration of Fig.~\ref{fig1}(b): 
(a) DOS for the interface Ti(0.5a,0)
next to the O-vacancy(0.5a,0.5a) in the TiO$_2$ layer; (b) the Ti(0.5a,0) densities of states projected on the rotated $3d$ orbitals.
Here the reconstructed $e_g$ orbital state corresponds to $3d_{3z^2-r^2}$ along the Ti-O$_v$-Ti and 
the $3d_{xz}$ corresponds to the old $3d_{xy}$ state.  (c) The densities of states for the interface Ti(0,0) more 
distant from the O-vacancy. The considered supercells comprise 4 uc thick LaAlO$_3$ layers and a 4 uc thick
SrTiO$_3$ layer.
}
\label{fig2}
\end{figure}

\subsection{Vacancy stripe configurations}

We start with the confiuration with the largest degree of clustering of O-vacancies in our models which is the vacancy stripes of type (b). 
Fig.~\ref{fig2} presents the spin polarized projected densities of $3d$ states for the Ti(0.5a,0)
atom nearest to the O-vacancies of type (b) and for the Ti(0,0) at the corners of the Ti$_2$O$_4$ plaquette in Fig.~\ref{fig1}(b) located
still in the interface layer but furthest away from the O vacancy stripe. 
For the Ti(0.5a,0) atom (Fig.~\ref{fig2}(a)), we find a strong ``splitting'' of 
the $3d_{3y^2-r^2}$ and $3d_{z^2-x^2}$ states with the $3d_{3y^2-r^2}$ being almost completely occupied with close to 2 electrons and the $3d_{z^2-x^2}$ almost completely empty and at very high energies above $E_f$. 
The extremely narrow $e_g$ DOS peaks 
are located in the energy window ($E_F-2.3$~eV;$E_F+0.5$~eV) 
relative to the Fermi level. This is in contrast to the DOS of the Ti(0,0) which is more distant to the oxygen vacancy
(Fig.~\ref{fig2}(c)). At  Ti(0,0) the empty $e_g$ states are located about 
$3.5$~eV above $E_f$ and 2~eV above the $t_{2g}$ states, the latter being partially occupied by the electrons generated due to the polar interface.

To better understand the nature of the sharp $e_g$ peaks, 
we plot in Fig.~\ref{fig2}(b) the $3d_{3y^2-r^2}$ density of states and see that actually the DOS in the sharp peaks is almost solely in an $e_g$ $3d_{3y^2-r^2}$ orbital with the lobes pointing along the Ti-O$_v$ direction, which implies a strong confinement of the vacancy-released electrons between the nearest Ti atoms. 
This strong preference of the $3d_{3y^2-r^2}$ orbital occupation is clearly seen also in the spatial charge density distribution in Fig~\ref{fig4}(b). 
The very sharp peaks in the $e_g$ density of states is a result of the quasi-one-dimensional character of the band structure of these mainly $3d_{3y^2-r^2}$ composed bands. 
For charge neutrality we need to accomodate two electrons per Ti in the dimerized chain which aside from the polar induced charge indicates that the Ti ions in the chains are formally divalent with nearly two electrons each. We see indeed from Fig.~\ref{fig2}(b) that both the spin up and down $3d_{3y^2-r^2}$ states are nearly fully occupied.These states are therefore rather magnetically inert. The empty $3d_{3y^2-r^2}$ stripe projected DOS,
also clearly visible in Fig.~\ref{fig2}(b), results from the other Ti ions which do not neighbor oxygen vacancies.   

The extreme splitting of the $e_g$ states $3d_{3y^2-r^2}$ directed along the Ti-vacancy direction reflects the
interface orbital reconstruction, a phenomenon recently observed at the interfaces between 
(Y,Ca)Ba$_2$Cu$_3$O$_7$ and La$_{0.67}$Ca$_{0.33}$MnO$_3$ \cite{chakhalian} and
predicted for interfaces between YBa$_2$Cu$_3$O$_6$ and STO \cite{pavlenko5,pavlenko2,pavlenko3}. 
For the oxygen deficient LAO/STO interfaces, the interface orbital reconstruction 
results in the change of the chemical valence state of the Ti atoms close to the O-vacancies. For high enough vacancy concentrations this strong valence change might  be visible in experiments like those of Ref.~\onlinecite{salluzzo}. 
Moreover, the conclusion of Ref.~~\onlinecite{salluzzo} about the in-plane mixed $3d_{xy}/3d_{x^2-y^2}$ character of the interface Ti electrons strongly supports the notion of the splitting of $e_g$ states and partial occupation of the in-plane vacancy-directed $e_g$ orbital found in our calculations.  

\begin{figure}[htbp]
\epsfxsize=8.5cm {\epsfclipon\epsffile{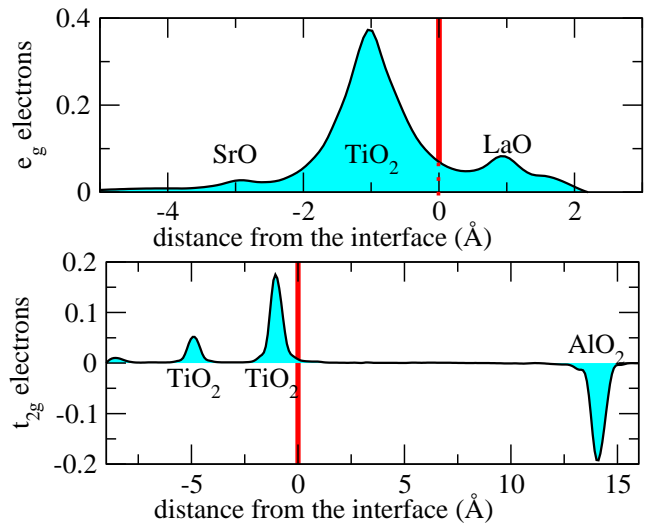}}
\caption{Orbital-projected charge density profiles along the [001] direction. Here the
O-vacancy of type (b) is located in the TiO$_2$ interfacial layer
in the supercell containing 4 uc thick LaAlO$_3$ layers and a 4 uc thick
SrTiO$_3$ layer. Note the large scale change between the upper and lower panels.
}
\label{fig3}
\end{figure}

To estimate the chemical valence of the Ti atoms, we have calculated the 2D electron charge density 
in two different energy windows below the Fermi level corresponding to occupied $e_g$ and $t_{2g}$
orbitals, and performed the integration in ($x$,$y$) planes across the interface. 
The results are presented in Fig.~\ref{fig3}. The integration of the $e_g$ charge profile
along the $z$-axis gives a total of about $1.9$ electrons, which implies that the excess vacancy-produced 
charge is almost fully transferred to the $e_g$ orbitals and induces a change of the chemical valence
of the nearest Ti atoms by $-1.9$. In contrast to the stoichiometric STO, the local $e_g$-states 
originating from the vacancies in the interface TiO$_2$ layers of the STO are placed well below the Fermi level.
The clustering of the vacancies into stripes at the TiO$_2$/LaO interface 
leads to a strong concentration of the vacancy-released electrons 
due to the shift of the filled $e_g$ subbands by more than 1~eV below the Fermi level. 
Integration of the conducting ($t_{2g}$) charge profile shown in 
Fig.~\ref{fig3}
reveals that the charge of the 0.3~$t_{2g}$ electrons in the TiO$_2$ layers near the
interface is compensated by the charge of the holes present at the AlO$_2$ surface of the LAO film.

\begin{figure}[htbp]
\epsfxsize=6.8cm {\epsfclipon\epsffile{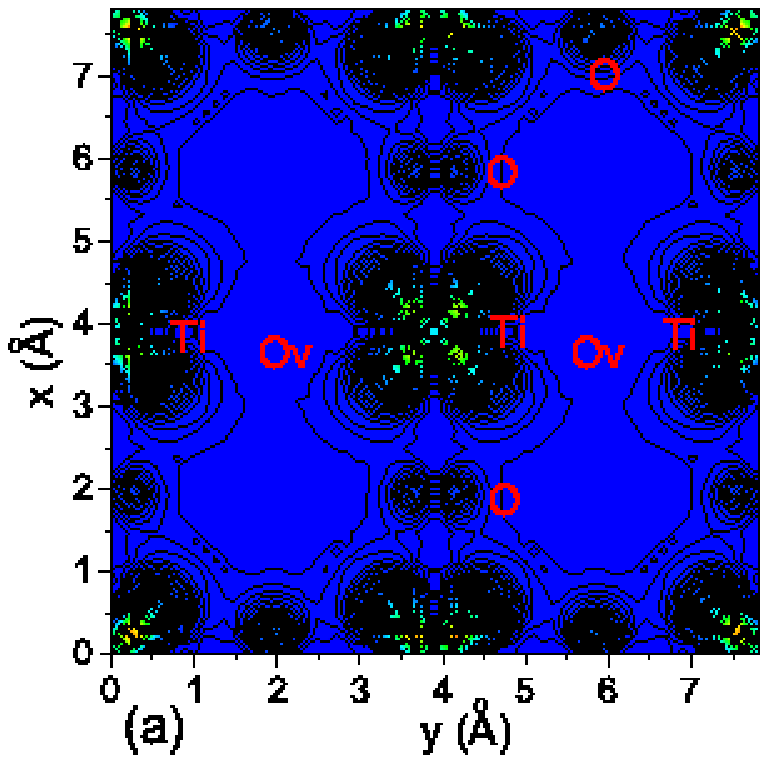}}
\epsfxsize=8.0cm {\epsfclipon\epsffile{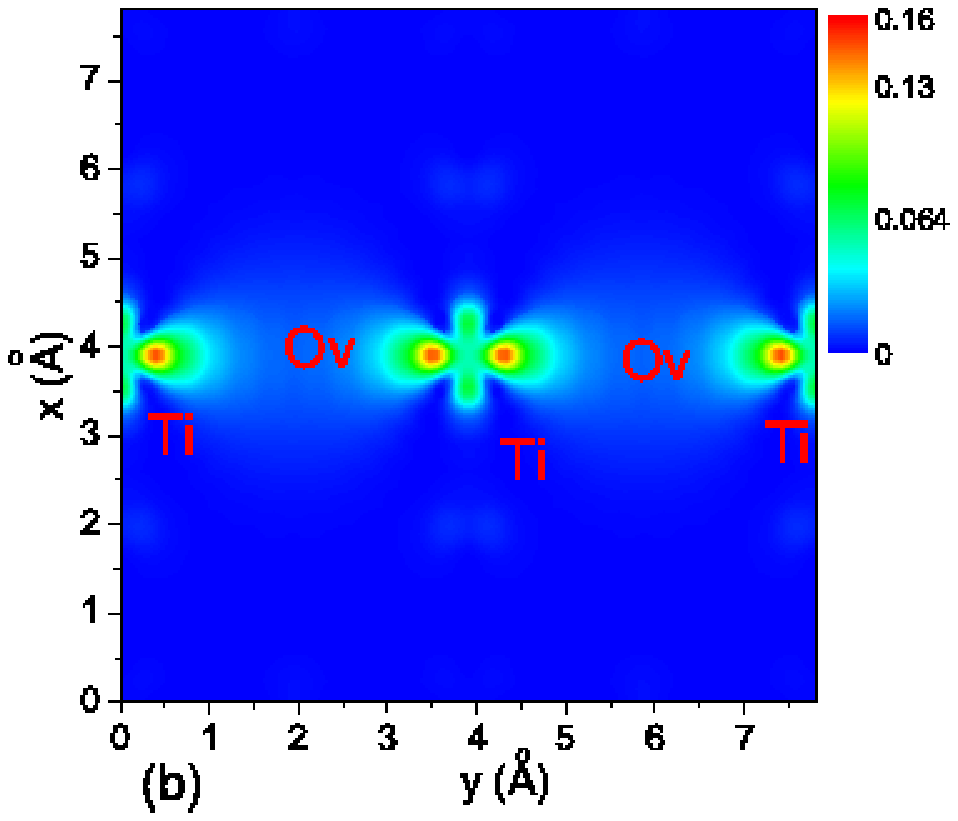}}
\caption{Electron density maps for vacancy stripes as shown in Fig.~1(b).  
(a) $t_{2g}$ and (b) $e_g$ electron states
for the interface TiO$_2$ layer with the O(0.5a,0.5b)-vacancy
in the supercell containing 4 uc thick LaAlO$_3$ layers and a 4 uc thick
SrTiO$_3$ layer. The contour lines in (a) indicate the $pd$-hybridization 
of the polar interface-induced electrons in TiO$_2$ layer.  
}
\label{fig4}
\end{figure}

In contrast to the strong spin polarization of the $t_{2g}$ states observed in Fig.~\ref{fig2}, the occupied $e_g$ states are almost unpolarized. 
The integration of the sharp $e_g$ peaks below the Fermi level gives a negligibly small
value of the spin polarization of about $0.001$~$\mu_B$, which implies the existence of 
the orbital separation of the spin
and charge degrees of freedom in the vacancy stripe configurations: the vacancy-released charge is localized
in the spin-nonpolar $e_g$-orbitals, whereas only the $t_{2g}$ polarity-induced intrinsic charge 
contributes to the magnetization. The contribution from different orbitals to the magnetization of various vacancy configurations
is presented in Table~\ref{tab1}. 

Fig.~\ref{fig4} shows the spatial distribution of the charge density generated in the energy windows below 
the Fermi level for the filled  $t_{2g}$ (Fig.~\ref{fig4}(a)) and $e_g$ states (Fig.~\ref{fig4}(b)).
One can see that the polarization-generated $t_{2g}$ electrons mostly occupy the Ti $3d_{xy}$ and
O $2p_y$ orbitals which are hybridized and do not contain any contribution from the oxygen vacancy state. 
In contrast, the vacancy-generated electronic charge in Fig.~\ref{fig4}(b) has a mixed 
$3d_{x^2-y^2}$/$3d_{3z^2-r^2}$ character 
(or $3d_{3y^2-r^2}$ character in the projection along the Ti-O$_v$-Ti).

Due to Coulomb attraction with the effective $2+$ charge of the nearest oxygen vacancy, 
the excess electron charge is trapped in the in neighboring Ti sites and because 
of the change in local $e_g$ -O 2p hybridization the charge is mainly found in Ti orbitals of $3d_{3y^2-r^2}$ character. 
These are placed around the center of the O-vacancy, similar to the bulk STO with point defects \cite{ricci}.

\subsection{Ti dimer-type O-vacancies in ($\sqrt{2} \times \sqrt{2}$)-plaquettes}

\begin{figure}[htbp]
\epsfxsize=8.0cm {\epsfclipon\epsffile{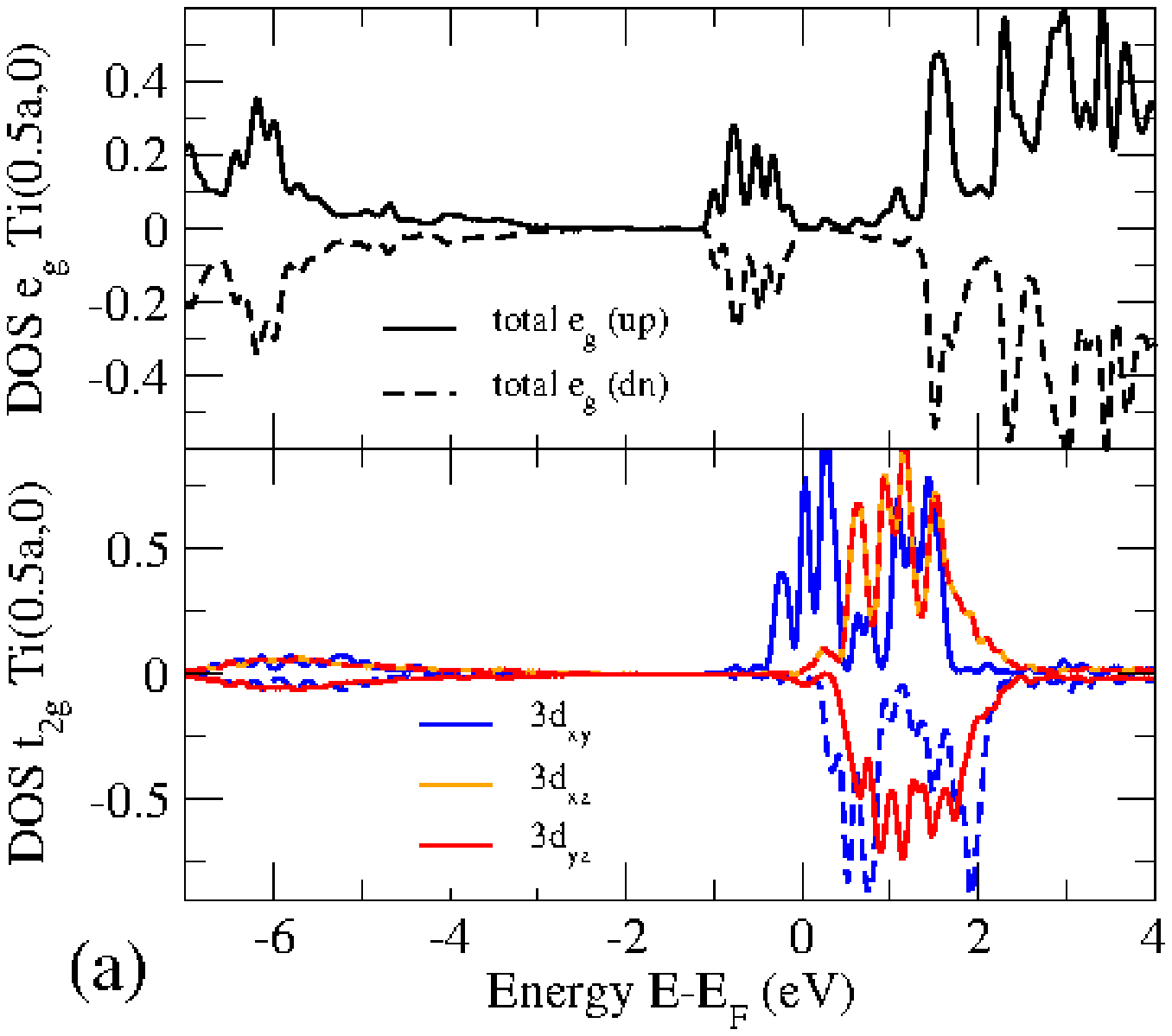}}
\epsfxsize=8.0cm {\epsfclipon\epsffile{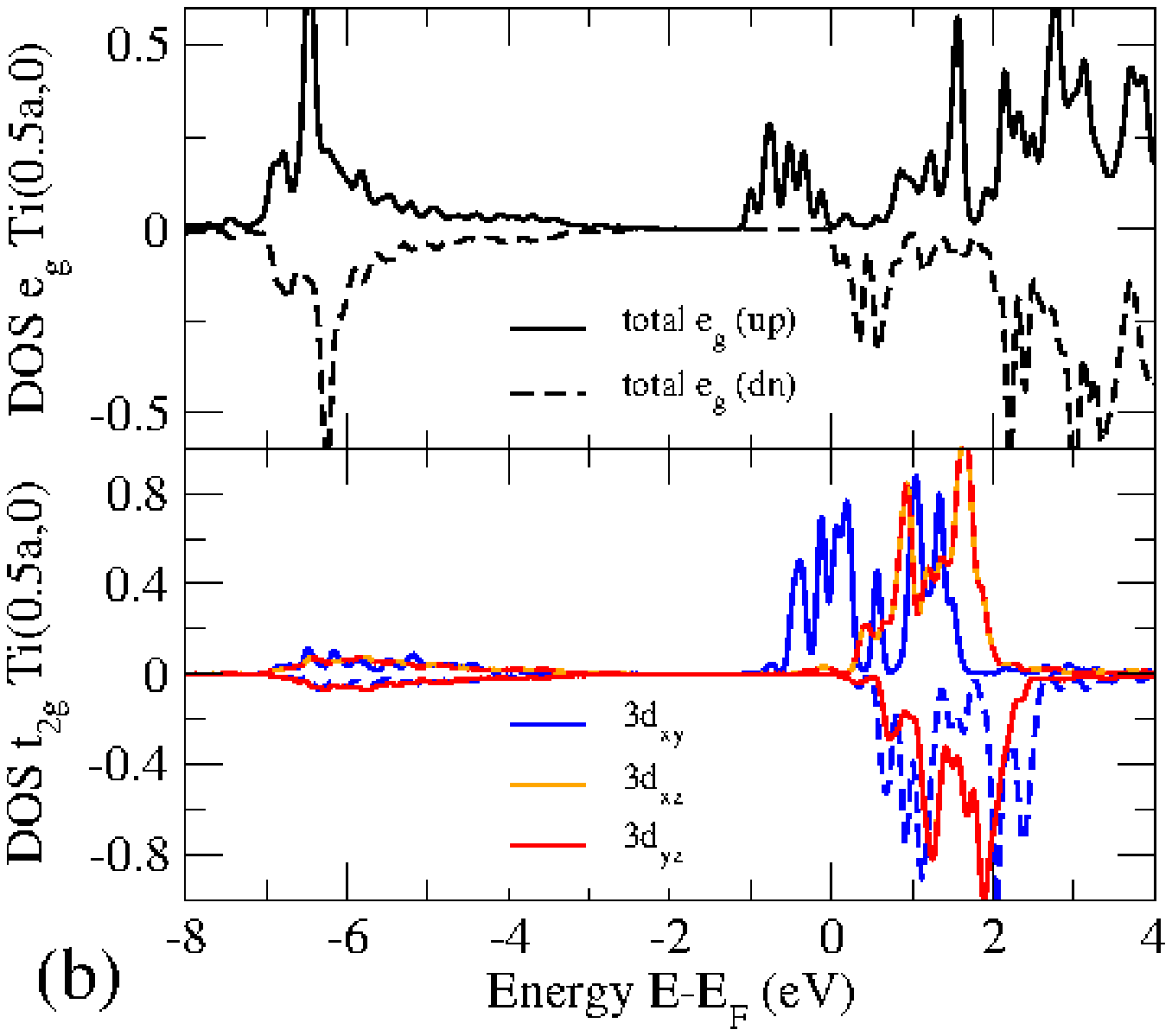}}
\caption{Spin-polarized orbital-projected densities of states for the interface Ti
in the interface TiO$_2$ layer with dimerized configurations (a) of oxygen vacancies.
Here the top panel (a) represents the DOS of a structure with unrelaxed Ti atoms fixed in the center and
in the corners of the Ti$_2$O$_4$ plaquette; the bottom panel (b) presents the $3d$ states of the relaxed structure with
the Ti atoms shifted by $0.05$~\AA\, outward from the O-vacancies which corresponds to the elongation of a vacancy-containing Ti-O$_v$-Ti dimer by $0.1$~\AA.  
}
\label{fig5}
\end{figure}

The strong localization of the excess electrons is already noted in the oxygen-reduced bulk STO, where 
the tendency for the formation of local Ti states is connected to the degree of clustering of the 
oxygen vacancies \cite{shanthi}. Similar to bulk STO, a weaker clustering 
in LAO/STO heterointerfaces reduces the energy gap between the filled and empty Ti $e_g$ states
to less than 1~eV,
which is demonstrated in Fig.~\ref{fig5} for the interface TiO$_2$ with Ti dimer-type O-vacancy configurations (a).

\begin{figure}[htbp]
\epsfxsize=7.5cm {\epsfclipon\epsffile{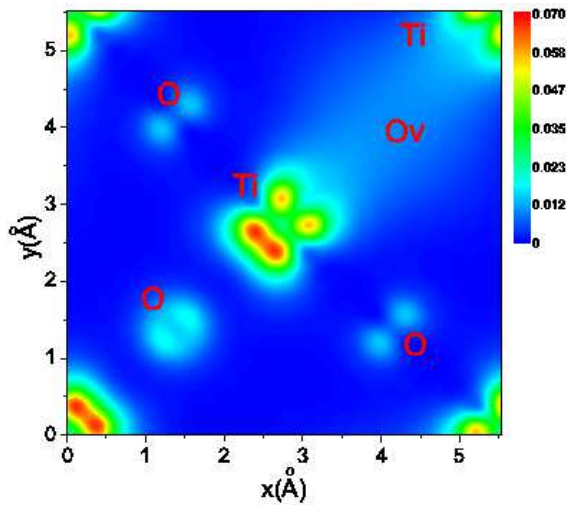}}
\caption{Charge density plot for an interface TiO$_2$ layer
with one O-vacancy at ($0.75a$, $0.75a$) ($a=5.523$~\AA) for each four O
in the Ti$_2$O$_4$ plaquette
in the supercell that containins two 4 uc thick LaAlO$_3$ layers and one 4 uc thick
SrTiO$_3$ layer. The plot has been obtained by calculating the electron densities in 
the energy window ($E_F-1.5$~eV; $E_F$) which corresponds to the mixed $e_g$+$t_{2g}$ states of Ti.
}
\label{fig6}
\end{figure}

\begin{figure}[htbp]
\epsfxsize=8.0cm {\epsfclipon\epsffile{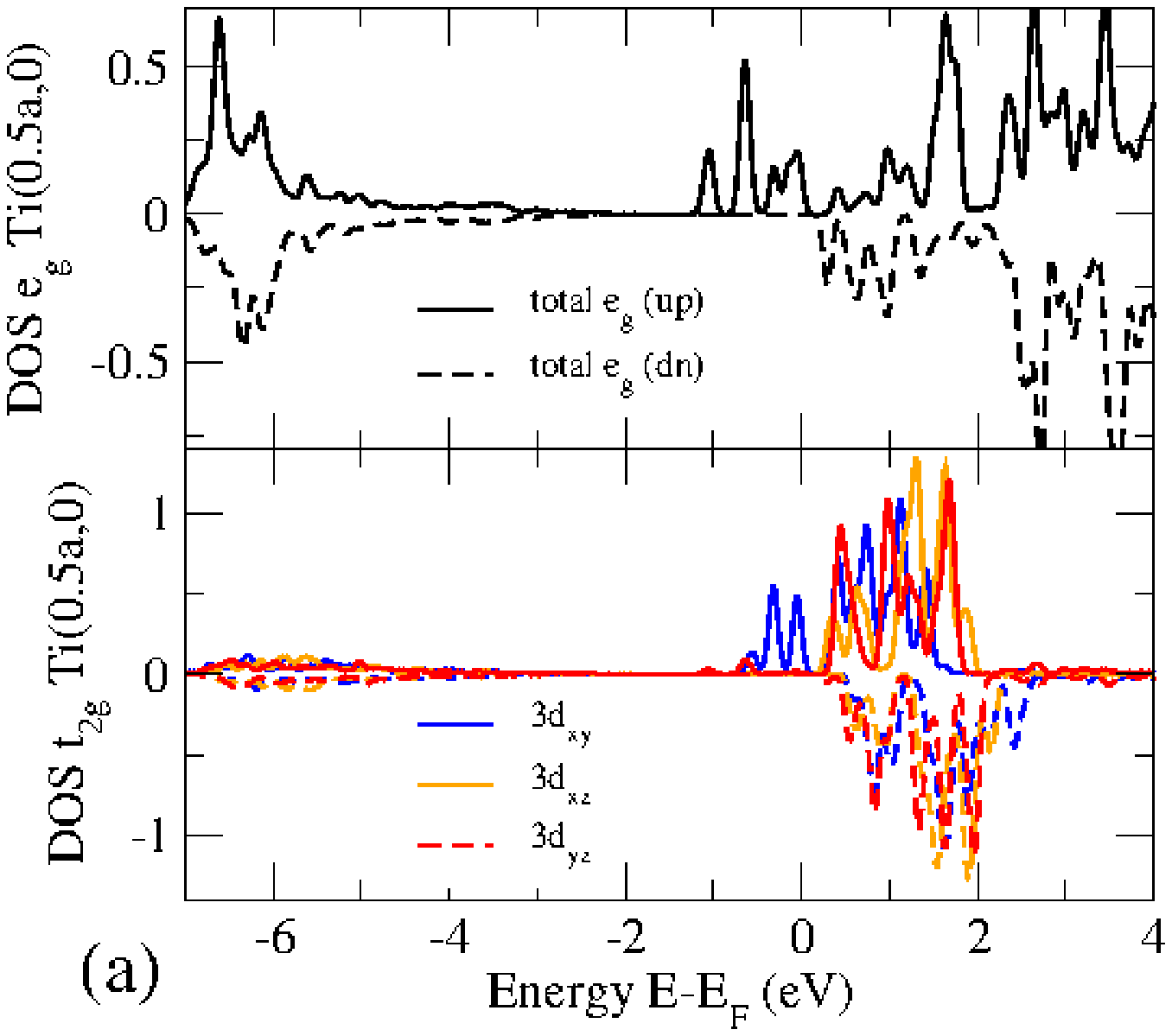}}
\epsfxsize=8.0cm {\epsfclipon\epsffile{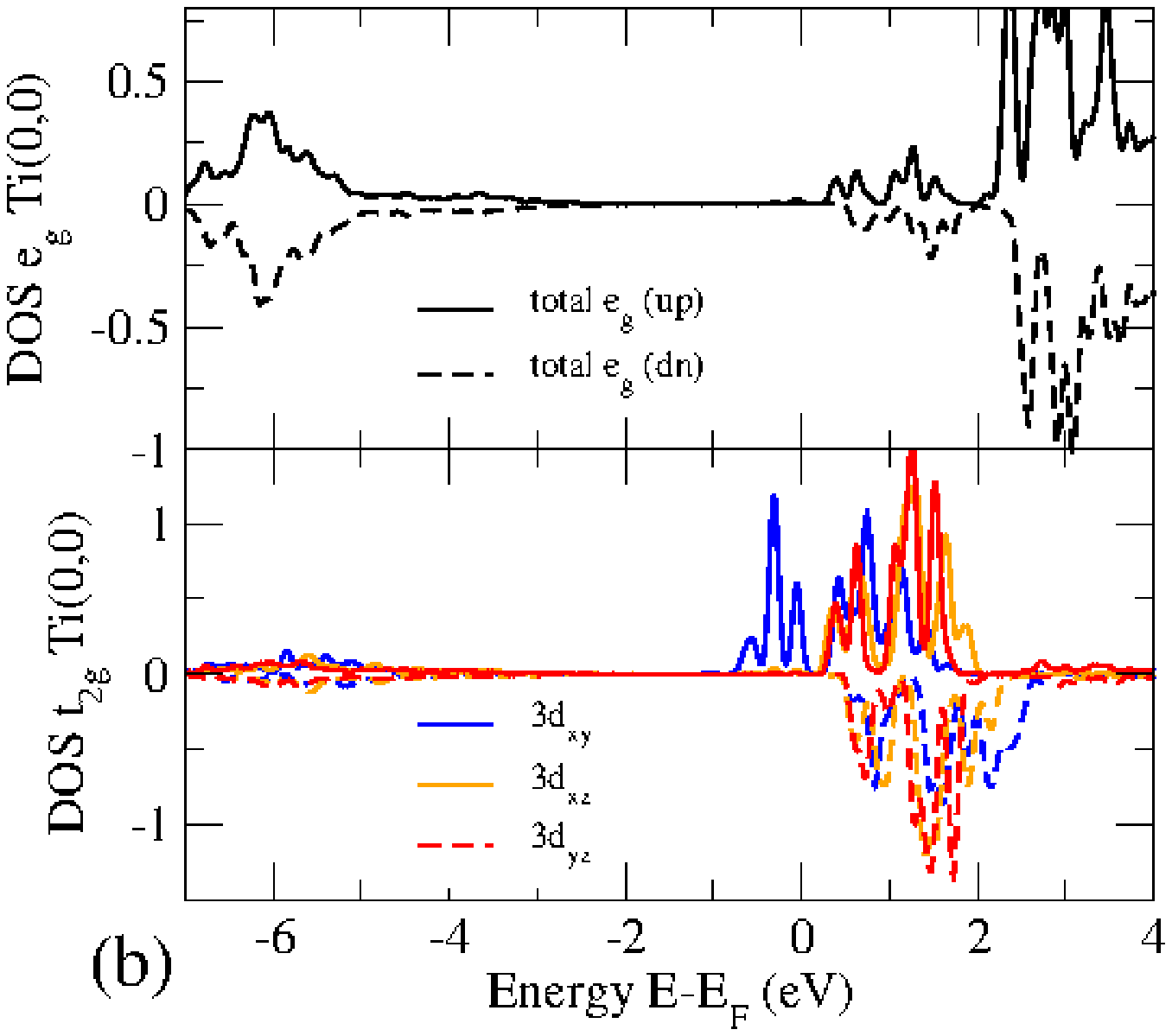}}
\caption{Spin-polarized orbital-projected densities of states for the interface Ti
in the interface TiO$_2$ $2\times 2$-plaquette configuration with 1/8-concentration of oxygen vacancies of tpe (c).
The positions of Ti, Al, La and  O atoms have been fully
relaxed in the ($x$, $y$)-planes and in the $z$-direction. 
The top (a) and bottom (b) plots represent the DOS for the Ti atoms nearest 
to the O-vacancy and the more distant corner Ti atom of the Ti$_4$O$_8$-plaquette.}
\label{fig7}
\end{figure}

In Fig.~\ref{fig5}a, the filled midgap $e_g$ states in the energy window ($E_F-1.2$~eV; $E_F-0.1$~eV) are 
strongly confined on Ti $e_g$ orbitals centered around the vacancies although
the occupied $e_g$ subband is broad as compared to the extremely narrow $e_g$ DOS-peaks 
in the striped configurations. The relaxation of the atomic positions leads to an elongation of the dimer
Ti-O$_v$-Ti by $0.1$~\AA\, due to the shifts of the Ti atoms by about $0.05$~\AA\, 
outward from the O-vacancy. The resulting DOS for the relaxed structure is presented in 
Fig.~\ref{fig5}b. The comparison with the DOS for the unrelaxed structure (top panel) shows 
a crucial effect of the structural relaxation of atoms around the O-vacancy. The relaxation 
leads to an additional magnetic splitting of the Ti $e_g$ states and to a decrease of the gap 
between the electronically occupied and the empty spin-up (majority) $e_g$ states. 
The occupied $e_g$ states are located in the energy window ($E_F-1.5$~eV; $E_F$) and hybridize 
with the $t_{2g}$ states. As a consequence, the electronic state of the Ti-O$_v$-Ti dimer is 
half-metallic with conducting electrons of mixed $e_g$-$t_{2g}$ character released in part 
by the O-vacancy and from the polar induced electronic reconstruction involving mainly 
$t_{2g}$ states. The mixed $e_g$-$d_{xy}$ character of the electrons in the Ti-O$_v$-Ti dimer 
reflects itself in the interface charge-density plot in Fig.~\ref{fig6}.
Due to the additional magnetic splitting of the $e_g$ states, the magnetic moment of Ti in 
the relaxed structure is strongly enhanced to $M_{\rm Ti}^{\rm rel}=0.475$~$\mu_B$, as compared 
to the unrelaxed structure 
with $M_{\rm Ti}^{\rm unrel}=0.15$~$\mu_B$ (see Table~\ref{tab1}).    

\begin{figure}[htbp]
\epsfxsize=8.5cm {\epsfclipon\epsffile{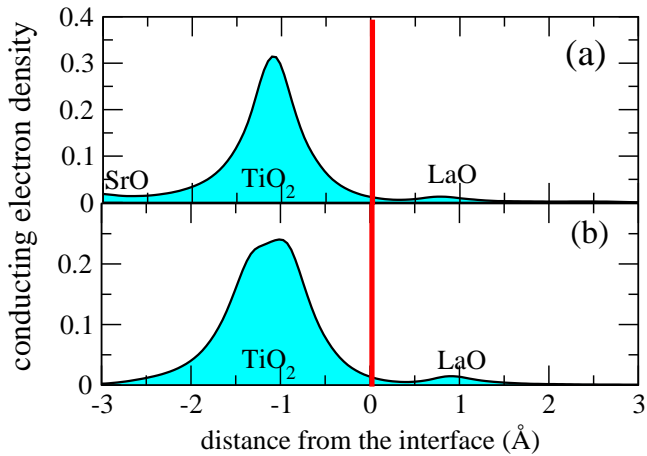}}
\caption{Profiles of the density of mobile charges along the [001] direction 
with $C_V=1/8$-concentration of oxygen vacancies of type (c).
The two profiles correspond to the O-vacancy placed (a) in the TiO$_2$ interface and (b) in the  
AlO$_2$ surface layer of ($2 \times 2$)-LAO/STO supercell.
}
\label{fig8}
\end{figure}

\begin{figure}[htbp]
\epsfxsize=8.5cm {\epsfclipon\epsffile{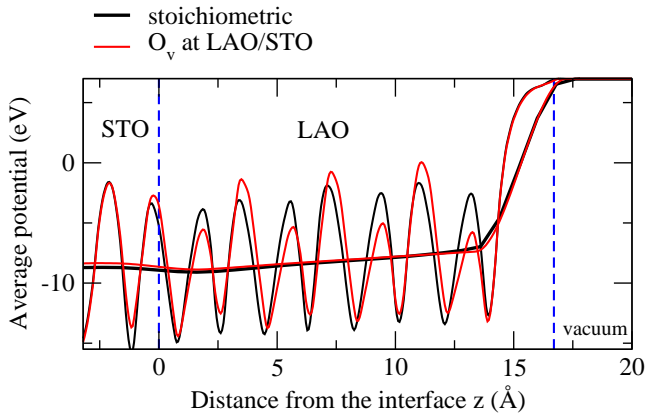}}
\caption{Profiles of $xy$-averaged electrostatic potentials along the [001] direction 
with $c_V=1/8$-concentration of 
oxygen vacancies of type (c) {\it at the interface}.
The black and red profiles correspond to the stoichiometric structure and the structure with 
the O-vacancy in the TiO$_2$ interface of ($2 \times 2$)-LAO/STO supercell, respectively.
}
\label{fig9}
\end{figure}

\begin{table}[b]
\caption{\label{tab1} Calculated magnetic moments of Ti ions nearest to the O-vacancy
and spin polarization character for (LaAlO$_3$)$_4$/(SrTiO$_3$)$_4$ heterostructures 
with different configurational state $n-(i)$ of O-vac. Here $n$ is the index of the MO$_2$-layer in 
STO or LAO with top surface/interface layer corresponding to $n=0$, and $i=a,b,c$ is the type
of vacancy configuration.
\\}

\begin{ruledtabular}
\begin{tabular}{llllllll} 
$n-(i)$  & character of polarization & $m_{\rm Ti}$($\mu_B$)  \\
\hline
0-(a) (STO)-relaxed & $t_{2g}+e_g$ & 0.475\\
0-(a) (STO)-unrelaxed & $t_{2g}$ & 0.15 \\
0-(b) (STO) & $t_{2g}$ & 0.34   \\
0-(c) (STO) & $t_{2g}$+$e_{g}$ & 0.47 \\ 
1-(b) (STO) & $t_{2g}$ & 0.12 \\
2-(b) (STO) & $t_{2g}$ & 0.07 \\
3-(b) (STO) & $t_{2g}$ & 0.006 \\
0-(b) (LAO) & $t_{2g}$ & 0.22 \\
0-(c) (LAO) & $t_{2g}$+$e_{g}$ & 0.56 
\end{tabular}
\end{ruledtabular}
\end{table}

\subsection{Ti dimer-type O-vacancies in ($2 \times 2$)-plaquettes}

To analyze lower concentrations of oxygen vacancies, we studied larger supercells which contain ($2 \times 2$)
M$_4$O$_8$-plaquettes in the ($x$,$y$) planes (Fig.~\ref{fig1}(c)). The elimination of one oxygen atom
in the interface (TiO$_2$)$_4$-plaquette corresponds to $c_V=1/8$-concentration or $1.5\cdot 10^{14}$~cm$^{-2}$ density 
of vacancies homogeneously distributed in the interface TiO$_2$ layer. 
The orbital projected Ti $3d$ DOS for the vacancy configuration of type (c) is displayed in Fig.~\ref{fig7}. 
As shown, even the 
smaller concentration of oxygen vacancies still causes a splitting of the $e_g$ orbitals of Ti atoms in the
close proximity of the vacancy. Similar to the other dimerized configurations, 
the occupied $e_g$ and $t_{2g}$ states are located in the same
energy window ($E_F-1.5$~eV; $E_F$) which implies a mixed $e_g$-$t_{2g}$ character of the vacancy-generated
electron states. Fig.~\ref{fig8}(a) presents the conducting electron density profile across 
the interface calculated 
by integrating the planar charge density generated in the energy range ($E_F-1.5$~eV; $E_F$). The location 
of the oxygen vacancies in STO leads to an insulating state of the AlO$_2$-surface. 
In Fig.~\ref{fig8}, the density plots are restricted to the interface region and are zero beyond 3\AA\, from the interface due to insulating character of the top AlO$_2$ layer. 
Moreover, the integration of the charge profile in Fig.~\ref{fig8}(a) 
gives exactly two electronic charges, in accordance with the charge released by the oxygen 
vacancy, which implies the absence of the polarity-induced interface electrons and a
suppression of the polar character of the LAO/STO interface. 
The reason for this at first glance surprising effect is discussed in the next section. 

\subsection{Vacancy-enhanced LAO-critical thickness}

\begin{table}[b]
\caption{\label{tab2} Atomic displacements (in \AA) in the interface TiO$_2$
and different (LaO)$_n$ ($n=1,\ldots,4$) planes of STO/LAO with
one O-vacancy of type (c) in the interface TiO$_2$ ($c_V=1/8$). 
Here the buckling of each layer is determined by the maximal atomic distortions 
in the corresponding MO-layer (M=Ti,La) which are defined as 
$\Delta z_{\rm MO}=\Delta z_{\rm M}-\Delta z_{\rm O}$.
The values $\Delta_0$ correspond to the buckling in the stoichiometric vacancy-free structure.
\\}
\begin{ruledtabular}
\begin{tabular}{llll}
(MO$_2$)$_n$  & $\Delta z_{\rm MO}$ & $\Delta_0 z_{\rm MO}$ & $\Delta-\Delta_0$ \\
\hline
(TiO$_2$)$_1$ & 0.21 & -0.1 & 0.31 \\
(LaO)$_1$ & 0.33 & 0.03 & 0.3 \\
(LaO)$_2$ & 0.28 & 0.05 & 0.23 \\
(LaO)$_3$ & 0.28 & 0.07 & 0.21 \\
(LaO)$_4$ & 0.36 & 0.08 & 0.32 
\end{tabular}
\end{ruledtabular}
\end{table}

An examination of the atomic positions in the structures with interface
O-vacancies reveals a considerable increase of the atomic distortions near
the vacancy in both the interface TiO$_2$ and in the LaO layers, presented in Table~\ref{tab2}. These distortions 
originate from the repulsion between La ionic charges and the nearest positive charged O-vacancy. Table~\ref{tab2} 
demonstrates a drastic increase of the 
La-O and Ti-O buckling which is related to an elongation of the local LaO bonds by about $0.2-0.3$~\AA\, in the LAO layer. 
The distortions in the LaO-layers have
an extended character along the [001]-direction and are significant at the interface as well as in the surface LaO-layers, which 
induces additional dipole moments antiparallel to the 
LAO layer polarization $P_{\rm LAO}=-N_{\rm LAO}c_{\rm LAO}e/2=-7.578$~e\AA\, (here $N_{\rm LAO}c_{\rm LAO}/2=7.578$~\AA\, is 
the thickness of the LAO layer). The dipole moments contribute to the  
compensation of the polarization field and, as a result, prevent the electronic reconstruction on account of a compensating polar charge. 
In the structure with $c_V=1/8$ vacancy concentration 
in the interface TiO$_2$, the additional vacancy-induced dipole polarization is $\Delta P_{{\rm TiO}_2}\approx 0.9$~e\AA\, in 
the interface TiO$_2$ layer and $\Delta P_{\rm LaO}=0.86$~e\AA\, in the subsurface LaO layer. 
This leads to the compensation of $P_{\rm LAO}$ by the sum of antiparallel contributions (1) $\Delta P_{\rm LAO}({\rm stoich})$ 
originating from the distortions in the stoichiometric structure and (2) $\Delta P_{\rm vac}$ caused by the enhanced 
distortions in the LAO layers due to O-vacancies in the TiO$_2$ layer. The prevention of the electronic reconstruction 
and intrinsic doping due to enhanced distortions in the LAO-layer implies a vacancy-driven enhancement of the LAO critical 
thickness below which the polar problem can be solved by atomic displacements in the LAO layer \cite{thiel,pentcheva}. 

Fig.~\ref{fig9} presents the microscopic $xy$-averaged and macroscopically averaged electrostatic potentials calculated 
in the stoichiometric LAO/STO with 4~u.c. thick
LAO layer, and in the structure with $c_V=1/8$ oxygen vacancies in the interface TiO2-layer. 
The macroscopic potential is calculated by the macroscopic averaging procedure proposed in Ref.~\onlinecite{baldereschi}. 
As compared to the stoichiometric system with the macroscopic electric field $E_{st}=0.17$~eV/\AA\, screened by 
the polar charge, the macroscopic electric field in the structure with the interface vacancies is 
screened by the field $\Delta P_{\rm vac}$e generated by antipolar distortions and approaches the value $E_{vc}=0.13$~eV/\AA\,. 
For the obtained internal field $0.13$~eV/\AA\,, the critical LAO-thickess sufficient for the dielectric breakdown of LAO/STO with 
the STO band gap $3.2$~eV \cite{demkov} in the structure with the interface O-vacancies is enhanced up 
to 24.6~\AA\, which corresponds to a 7~u.c.~thick LAO layer.

\section{A model for the $e_g$ level splitting}

\begin{figure}[tbp]
\epsfxsize=8.0cm {\epsfclipon\epsffile{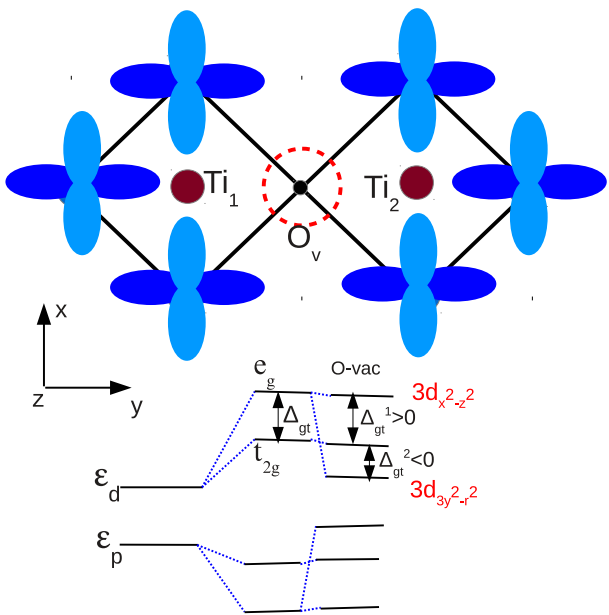}}
\caption{Schematic view of a Ti$_1$-O$_v$-Ti$_2$-cluster
that models an oxygen vacancy with the corresponding reconstruction of the $3d$ electron 
levels. Here the bare unrenormized energies $\varepsilon_p=-10.199$~eV and $\varepsilon_d=-6.01$~eV; 
$\Delta_{gt}=2.7$~eV, $\Delta_{gt}^1=3.2$~eV, and $\Delta_{gt}^2=-2$~eV.
}
\label{fig10}
\end{figure}

The vacancy-induced  Ti 3$d$ orbital reconstruction can be understood from the analysis of the local bonding 
for a two-electron state of a Ti-O$_v$-Ti-cluster (Fig.~\ref{fig10}).

We focus on the contribution of the covalent Ti$3d$-O$2p$ bonding to the diagonal and exchange 
contributions to the electron energy levels which determine the magnetic state 
(paramagnetic singlet or ferromagnetic triplet) of the two-electron cluster. 

In the stoichiometric SrTiO$_3$, the two-electron energy contains a $pd$-hybridization term
due to the overlap between the $3d$-states ot Ti atoms and $2p$-states of the central oxygen atom.
In a TiO$_6$-octahedron, such a covalent contribution $V_{pd}$ produces shifts 
of the one-electron $3d$ energy levels\citep{wolfram}

\begin{eqnarray} \label{pd_hyb}
\!\!\varepsilon(E)&& =\frac{1}{2}(\varepsilon_e+\varepsilon_{||})+
\frac{1}{2}\sqrt{(\varepsilon_e-\varepsilon_{||})^2+8V_{pd\sigma}^2\left(\sum_\alpha s_\alpha^2\pm s^2\right)}\nonumber\\
\!\!\varepsilon(T_2)&& =\frac{1}{2}(\varepsilon_t+\varepsilon_{\perp})+
\frac{1}{2}\sqrt{(\varepsilon_t-\varepsilon_{\perp})^2+16V_{pd\pi}^2\sum_{\alpha=x,y} s_\alpha}
\end{eqnarray}
Here $\varepsilon_t=-6.258$~eV and $\varepsilon_e=-5.638$~eV are the Ti $3d$ ionization energies 
plus Madelung potential, renormalized by the electrostatic shifts due to the cubic field. The corresponding
energies for the O $2p$ states are: $\varepsilon_{||}=-10.519$~eV and $\varepsilon_{\perp}=-10.039$~eV. In 
Eq.~(\ref{pd_hyb}), $s_{\alpha}=\cos k_{\alpha} a$ $(\alpha=x,y,z)$ and $s^2=\sqrt{\sum_{\alpha} s_{\alpha}^4-\sum_{\alpha 
\ne \beta} s_{\alpha}^2 s_{\beta}^2}$. With the $pd$-covalency parameters
$V_{pd\sigma}\approx 2.1$eV and $V_{pd\pi} \approx 0.84$eV defined in Refs.~\onlinecite{kahn,mattheiss}, 
we can estimate the splitting between the anti-bonding $e_g$- and $t_{2g}$-states
in the $\Gamma$-point of the Brillouin zone \cite{kahn,wolfram,mattheiss}:
\begin{eqnarray}
\Delta_{gt}=\varepsilon(E)-\varepsilon(T_2)\approx \frac{\tilde{V}_{pd\sigma}^2-\tilde{V}_{pd\pi}^2}{\Delta_0}\approx 2.7~{\rm eV},
\end{eqnarray}
where $\tilde{V}_{pd\sigma}=\sqrt{6}V_{pd\sigma}$, $\tilde{V}_{pd\pi}=2\sqrt{2}V_{pd\pi}$, and $\Delta_0\approx 4.5$~eV is 
the bare gap between the $2p$ and $3d$ energy levels, unrenormalized by the covalent overlap.
 
The elimination of the oxygens in a $\cdots$-O-Ti-O-Ti-$\cdots$ stripe along the $y$-direction is equivalent 
to the condition $V_{pd\sigma} 
s_y=V_{pd\pi} s_y=0$. Consequently, the absence of the local covalency lowers the local symmetry and leads
to a splitting of the $e_g$ states with the lower energy state corresponding to a $3d_{3y^2-r^2}$ orbital energy 
with lobes along the bond direction and the one at an almost unchanged energy of $3d_{x^2-z^2}$ i.e. with lobes 
in a plane perpendicular to the Ti-O$_v$ direction 
(see scheme of Ti-O$_v$-Ti cluster in Fig.~\ref{fig10}) with the splitting energies
\begin{eqnarray}
\Delta_{gt}^1 \simeq \frac{\tilde{V}_{pd\sigma}^2-\tilde{V}_{pd\pi}^2/2}{\Delta_0}\approx 3.2~{\rm eV},\\
\Delta_{gt}^2 \simeq -\frac{\tilde{V}_{pd\sigma}^2/3+\tilde{V}_{pd\pi}^2/2}{\Delta_0}\approx -2~{\rm eV}
\label{delta2}
\end{eqnarray}
These estimates are for the stripe configuration which two O-vacancy neighbors for each Ti. 
The negative splitting $\Delta_{gt}^2$ implies a strong negative shift of the $e_g$-orbital with lobes pointing to the vacancy. 
The $e_g$-orbital is shifted below the
$t_{2g}$ energy levels, the corresponding orbital inversion is observed in the orbital density of states in Fig.~\ref{fig2}(b). 
In the dimerized Ti-O$_v$-Ti configuration (a) shown in Fig.~\ref{fig10}, the smaller splitting energy $\Delta_{gt}^2\approx -1$~eV 
results from the missing of only half of oxygens in the vacancy stripe configuration.

\begin{figure}[tbp]
\epsfxsize=4.2cm {\epsfclipon\epsffile{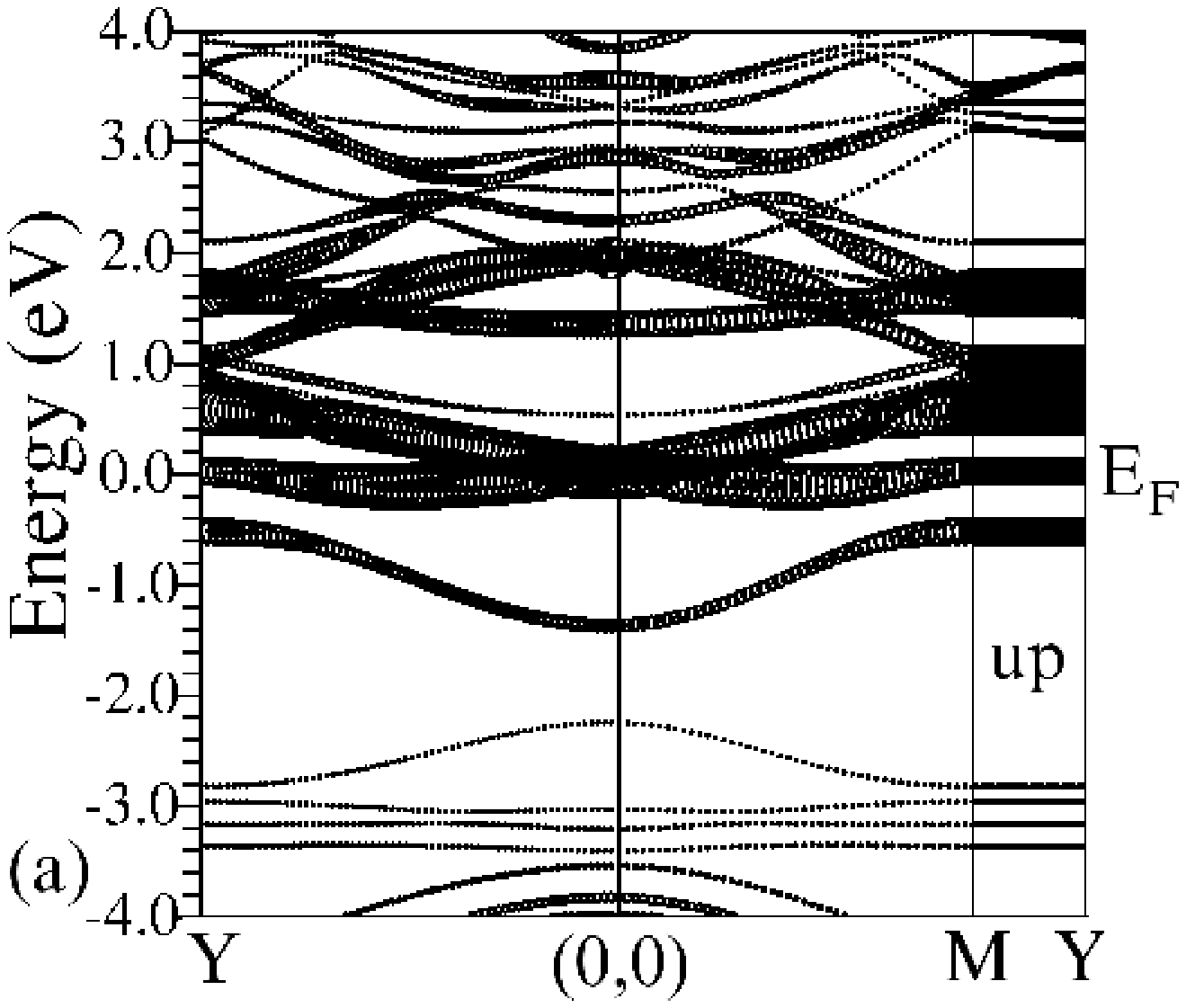}}
\epsfxsize=4.2cm {\epsfclipon\epsffile{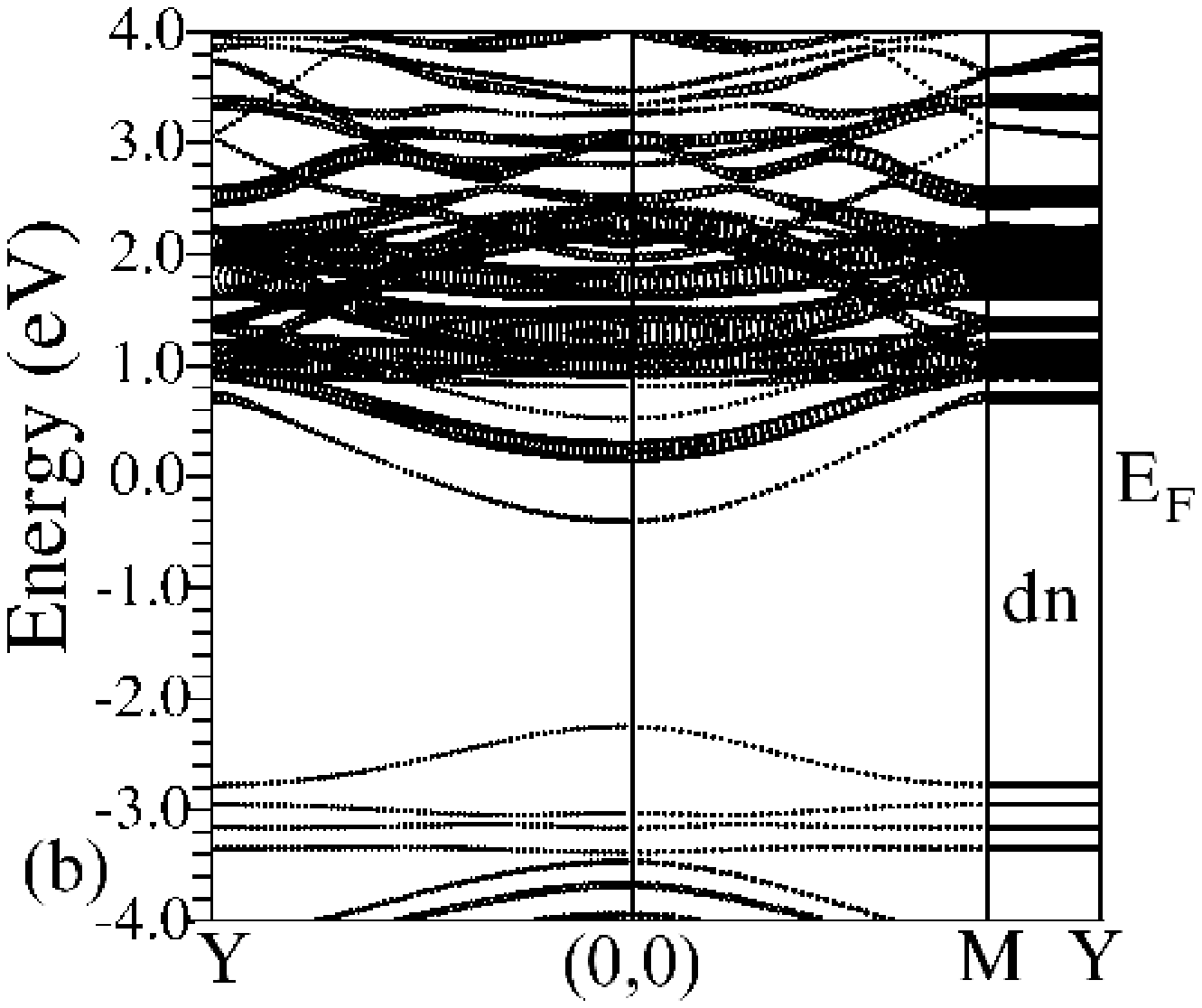}}
\epsfxsize=4.2cm {\epsfclipon\epsffile{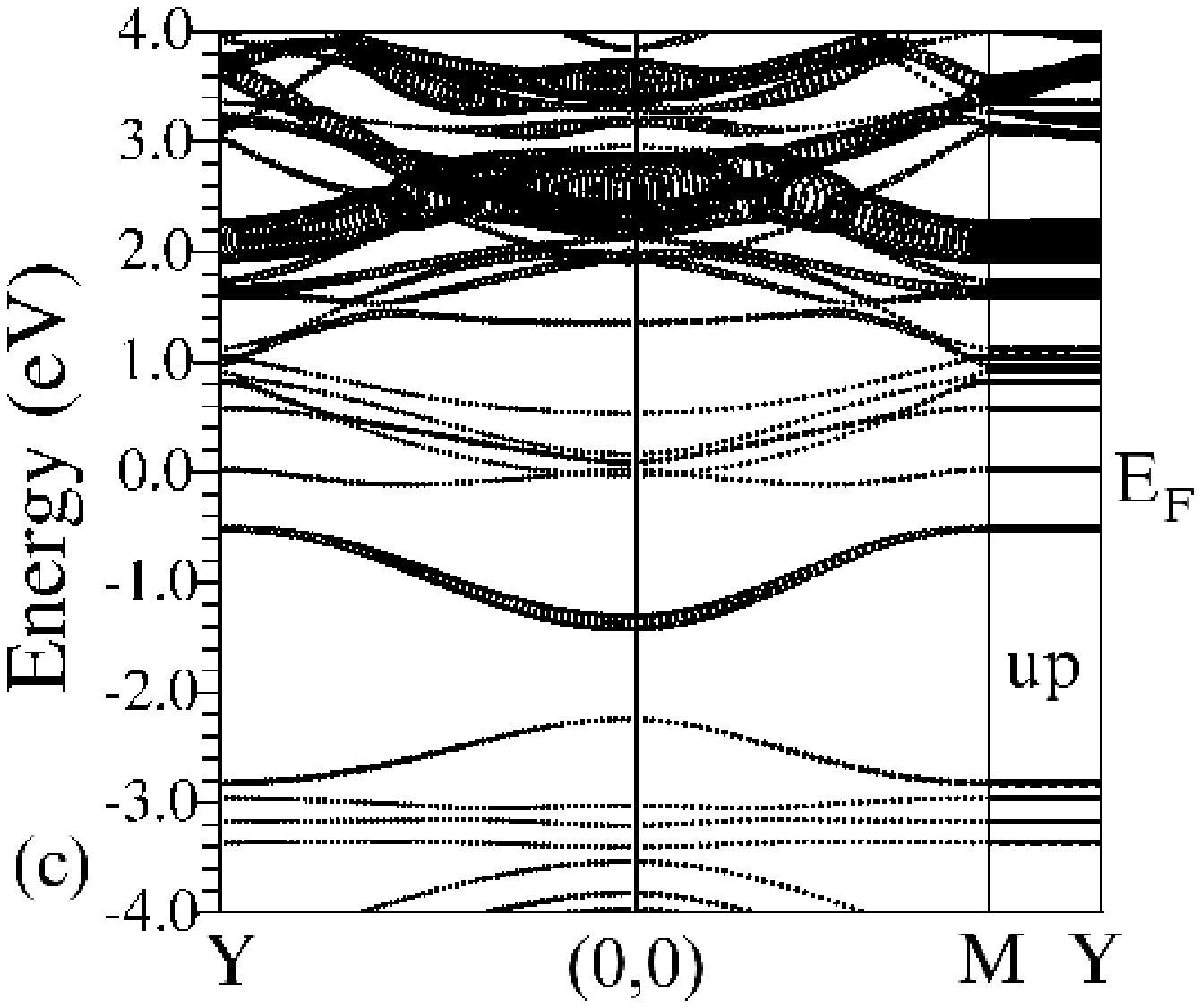}}
\epsfxsize=4.2cm {\epsfclipon\epsffile{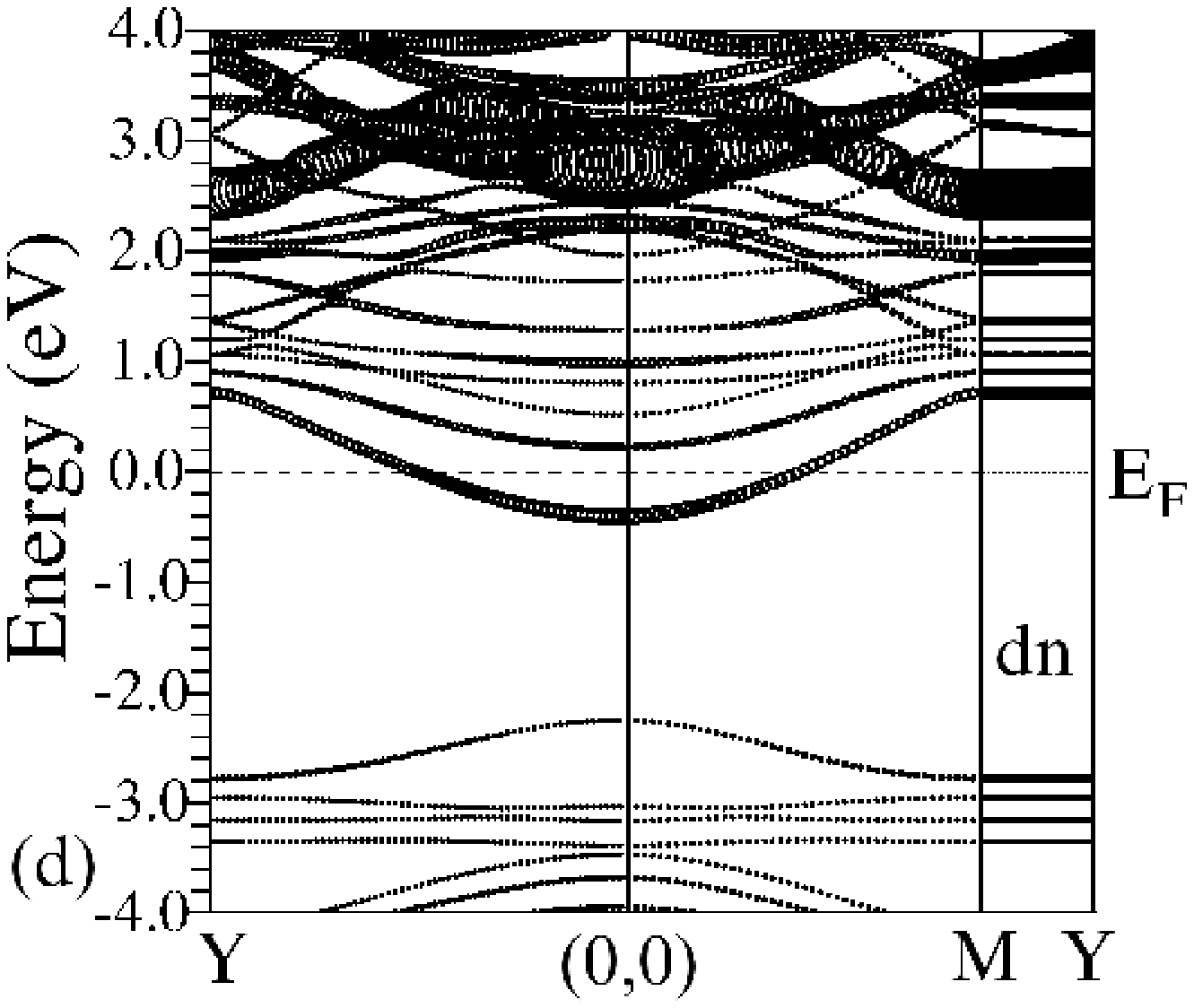}}
\caption{Band structure for a one uc thick SrTiO$_2$ with an oxygen vacancy
of type (a) in the surface TiO$_2$ layer. The thick lines indicate the bands with (a), (b) $3d_{x^2-y^2}$ character 
and (c), (d) $3d_{3z^2-r^2}$ character in the rotated coordinate system 
shown in Fig.~\ref{fig1}(a). Panels (a), (c) and panels (b), (d) display the up-spin and down-spin energies, respectively.}
\label{fig11}
\end{figure}

The vacancy-induced orbital reconstruction is not restricted to the LAO/STO interfaces, but has a generic character 
for the titanate surfaces and interfaces. It is related to the specific electronic structure of Ti, with the 
covalence-induced splitting between the high-energy $e_g$ and low-energy $t_{2g}$ states. For example, 
in a TiO$_2$-terminated (001)-surface layer of a SrTiO$_3$ slab, the absence of the top covalent-bonded 
oxygens results in the five-fold coordination of surface Ti atoms with consequent $d$-orbital splitting 
and reduction of the energy gap by about $0.6$~eV \cite{wolfram,padilla}. As a consequence, a surface O-vacancy 
in a TiO$_2$-terminated (001)-layer is equivalent to the additional condition $V_{pd\sigma}s_z=0$ in 
Eq.~(\ref{pd_hyb}), which leads to a negative shift of the energy of one of the surface $e_g$-states 
\begin{eqnarray}
\Delta_{gt}^2 \simeq -\frac{\tilde{V}_{pd\pi}^2/2}{\Delta_0}\approx -0.13~{\rm eV},
\end{eqnarray}
similarly to  $\Delta_{gt}^2$ of  Eq.~(\ref{delta2}) for an interface layer.
Our calculations of a one-uc thick STO-slab 
indeed confirm the formation of a surface $3d_{3y^2-r^2}$ state in the gap shifted by $0.6$~eV below 
the $3d_{xz}$ band (see Fig.~\ref{fig11}).  

\begin{figure}[b]
\epsfxsize=6.5cm {\epsfclipon\epsffile{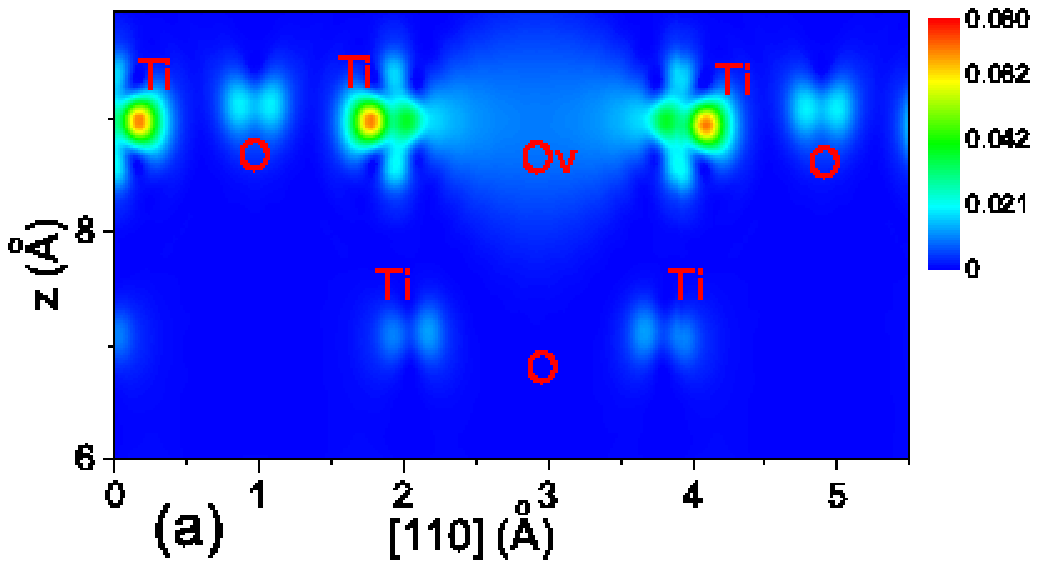}}
\epsfxsize=6.5cm {\epsfclipon\epsffile{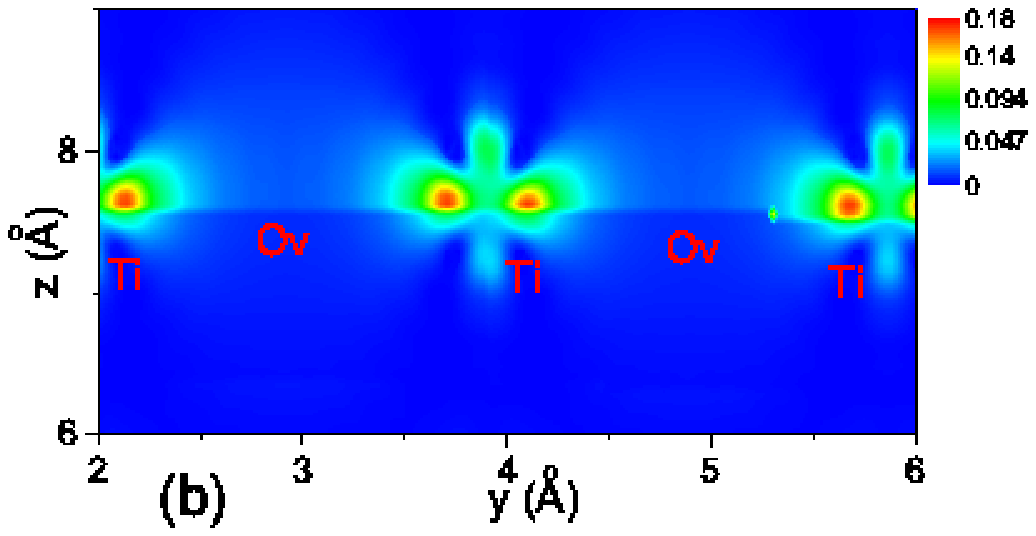}}
\caption{Charge density plots across the (001)-interface for the O-vacancies of dimer-type (a) and stripe-type (b) in the interface TiO$_2$ layer.
}
\label{fig12}
\end{figure}

In the bulk stoichiometric SrTiO$_3$, the nonmagnetic $3d$-electron state is associated with a small 
exchange energy splitting of about $0.06$~eV separating singlet and triplet states \cite{soules}. 
As follows from the GGA+U electron-density contours~(Figs.~\ref{fig4}(b) and \ref{fig6}), the oxygen vacancy induces a strong spatial shift of the 
$3d$ electron density between Ti atoms of about $0.5$~\AA\, towards the center of Ti-O$_v$-Ti-dimer, which
increases the overlap exchange integrals by about 0.36~eV at the LAO/STO-interfaces of type (b) (Fig.~\ref{fig12}). 
For the dimerized vacancy configuration (a), we obtain from the comparison of the total energy of ferromagnetic and antiferromagnetic configurations 
the value of the exchange of $J=0.28$~eV, however we note that the exchange energy is typically overestimated in the DFT calculations. 

The structural relaxation of the vacancy surroundings leads to alternating rotations of the $z$-directed Ti$_1$O and Ti$_2$O bonds 
by $\Delta\Theta=\pm 8^\circ$ and to a tilting of the reduced TiO$_5$-octahedra towards the vacancy center 
in the ($x$,$z$)- and ($y$,$z$)-planes (Fig.~\ref{fig11}), which increases the overlap. 
Due to the large $J>0$, 
the two-electron local state on a Ti-O$_v$-Ti forms a triplet with a triplet energy $E_2-J$ where $E_2$ is the two-electron energy in the Ti-O$_v$-Ti cluster. 
Due to the planar electron transfer between the nearest Ti$_2$O$_4$-plaquettes, the magnetic triplet state originating from the localized O-vacancies, 
is spread within the interface TiO$_2$-plane and is to be considered as two-dimensional magnetic ordering stabilized 
by the exchange splitting of the surface $3d$ bands.     

\begin{figure}[htbp]
\epsfxsize=7.5cm {\epsfclipon\epsffile{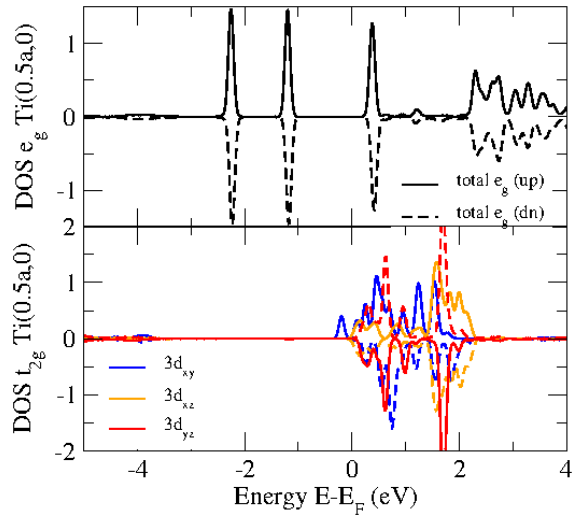}}
\caption{Spin-polarized orbital-projected densities of states for the Ti
near the O-vacancy(0.5a,0.5a) in the vacancy-stripes configuration (b) in the layer second from the interface TiO$_2$ plane.
For this calculation, a supercell LAO(4uc)/STO(4uc) has been considered.}
\label{fig13}
\end{figure}

The notion that the magnetic ordering at the interface has a two-dimensional character is confirmed by calculations 
of the supercells in which the oxygen vacancy is residing in more distant layers from the interface TiO$_2$-layer.
In spite of the occurrence of the local orbital reconstruction in the vicinity of the O-vacancy (Fig.~\ref{fig13}),
the local magnetic splitting of $3d$ states is weak and the magnetic moments of Ti in the layer with the vacancies 
are strongly reduced to values of 0.06-0.12~$\mu_B$ in the layer second from the interface TiO$_2$-layer, 
and to 0.006~$\mu_B$ in the layer fourth from the interface TiO$_2$. The location of the O-vacancies 
in the layers distant from the interface leads to an interlayer exchange between the 
vacancy-containing and interface TiO$_2$ layers, which induces a weak magnetic moment in the interface 
layer. This effect is similar to the double-exchange induced ferromagnetism between transition 
metal ions in different oxidation state\cite{zener}.
Vacancies in the TiO$_2$ layer second from the interface reduce the magnitude of the 
local Ti magnetic moments in the vacancy-free interface layer to $0.16$~~$\mu_B$, as compared 
to the magnetic moments in the range of 0.34--0.5~$\mu_B$ in heterostructures with interface vacancies.

\section{Oxygen vacancies at the LAO surface}

Another possible source of conducting interfacial charge are electrons generated by oxygen vacancies
in the LAO. Such vacancies are expected to occur predominantly in the top AlO$_2$ surface layer~\cite{zhang,pavlenko}.
To explore the electronic state of the oxygen vacancies in the AlO$_2$ surface, we consider first
a vacancy placed in the center of the ($2 \times 1$) Al$_2$O$_4$-plaquette with $c_V=1/4$ . 
In this case, the surface atomic 
configuration is described as Al$_2$O$_3$, and the charged LAO/STO supercell is expected to be doped by the two excess
electrons to preserve the overall electrostatic neutrality. To study the supercell charging, we
have determined the two-dimensional charge densities in the energy window ($E_F-1$~eV; $E_F$) across the interface
which corresponds to the charge occupation of the 3$d$-conducting bands.

\begin{figure}[bp]
\epsfxsize=8.5cm {\epsfclipon\epsffile{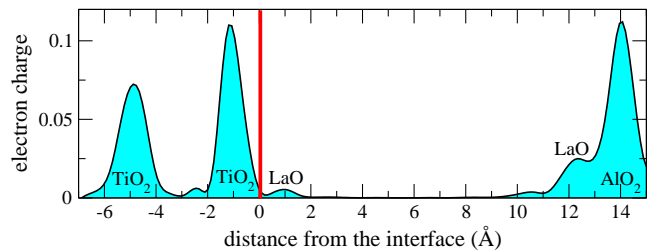}}
\caption{Charge density profiles along [001] direction for a
O-vacancy of type (b) located in the AlO$_2$ surface layer
in the supercell containing 4 uc thick LaAlO$_3$ layers and a 4 uc thick
SrTiO$_3$ layer.
}
\label{fig14}
\end{figure}

\begin{figure}[htbp]
\epsfxsize=7.5cm {\epsfclipon\epsffile{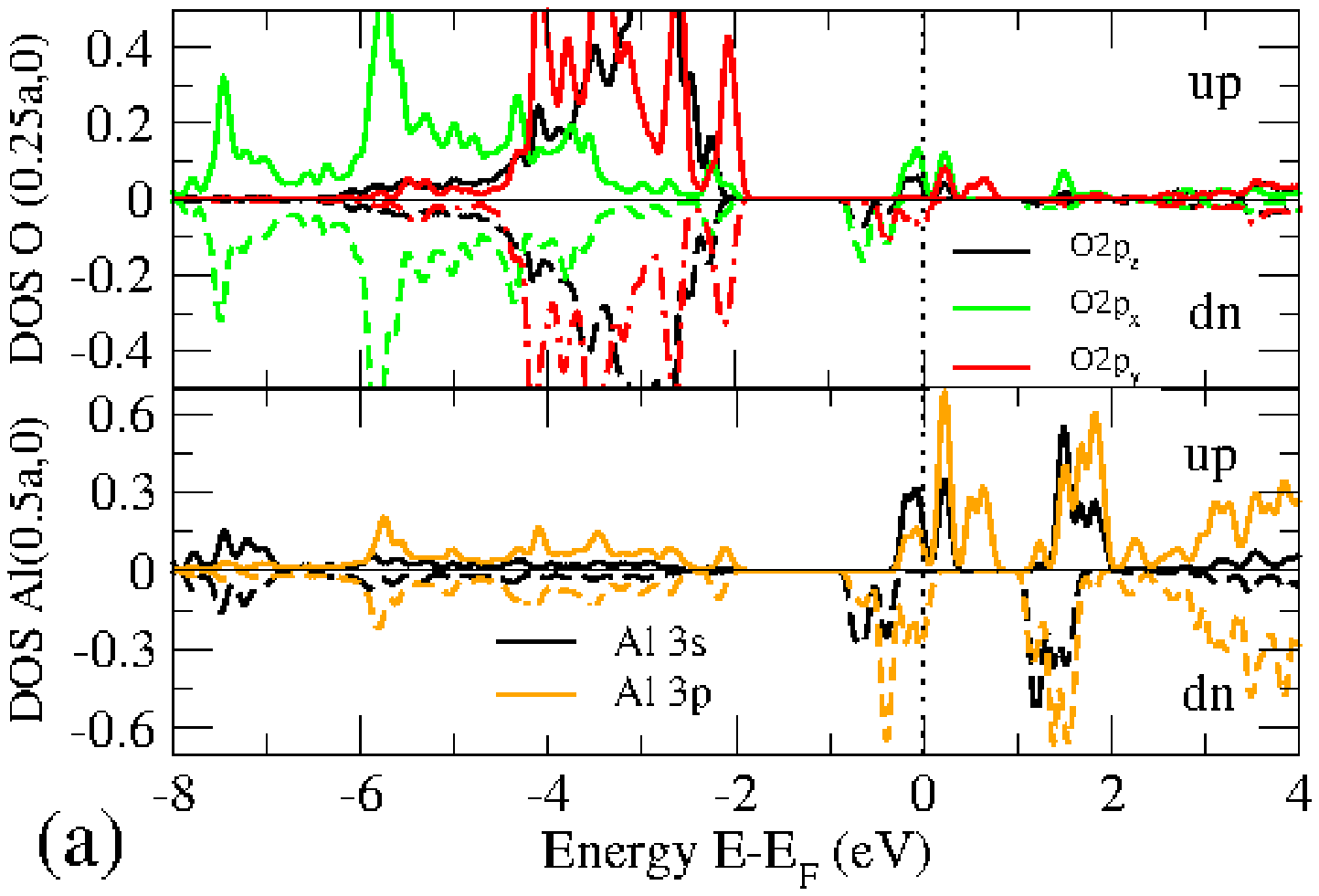}}
\epsfxsize=6.0cm {\epsfclipon\epsffile{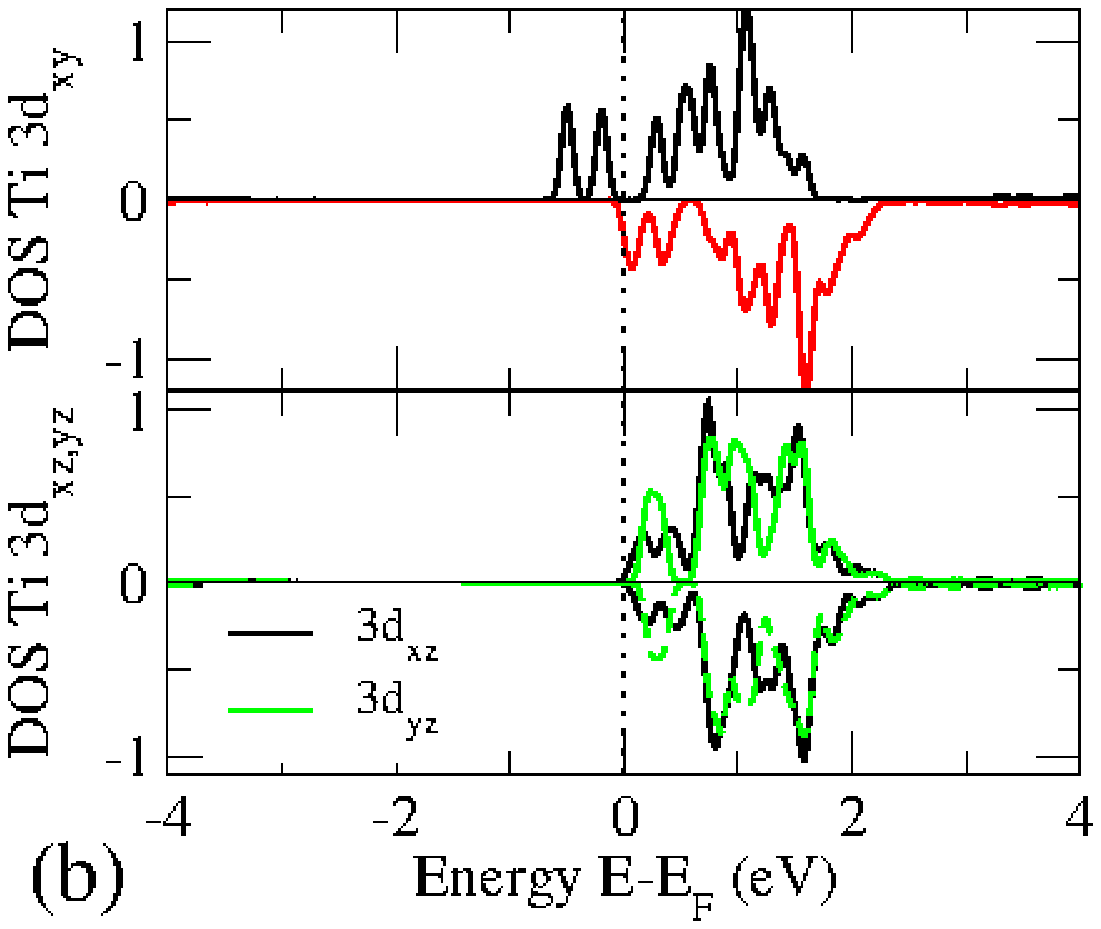}}
\caption{Projected DOS (a) for the surface AlO$_2$ and (b) interface TiO$_2$ layer
in LaAlO$_3$(4 uc)/SrTiO$_3$(4 uc) with one O-vacancy of type (b) being present in the AlO$_2$ surface.
}
\label{fig15}
\end{figure}

Fig.~\ref{fig14} presents the charge profile obtained by planar integration of the 
calculated electron densities. The integration yields two electron charges per 
interface unit cell, with $1.25$ of electron charges distributed in the surface and 
the remaining $0.75$ in the STO/LaO-interface layer. We note the absence of polar charge 
produced by polar discontinuities in the vacancy-free LAO/STO. Consequently, the surface oxygen 
vacancies suppress the polar field of LAO due to the supply of the polarity-compensating 
excess electrons 
between the surface and interface layers. As a result, the interface charging 
in LAO/STO with surface vacancies has a self-doping character.

\begin{figure}[htbp]
\epsfxsize=6.8cm {\epsfclipon\epsffile{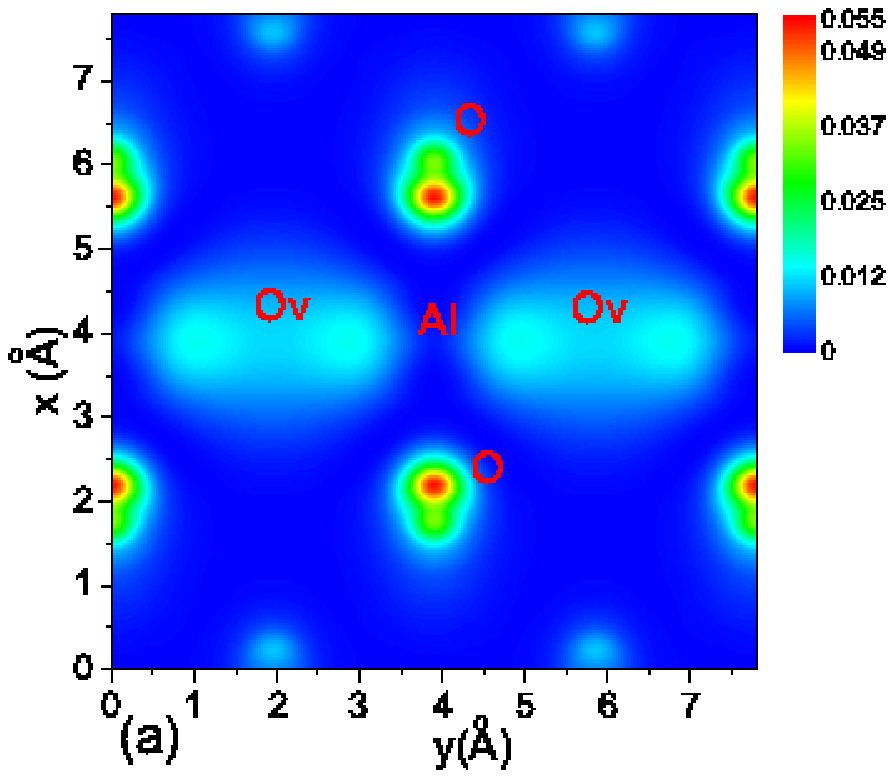}}
\epsfxsize=6.5cm {\epsfclipon\epsffile{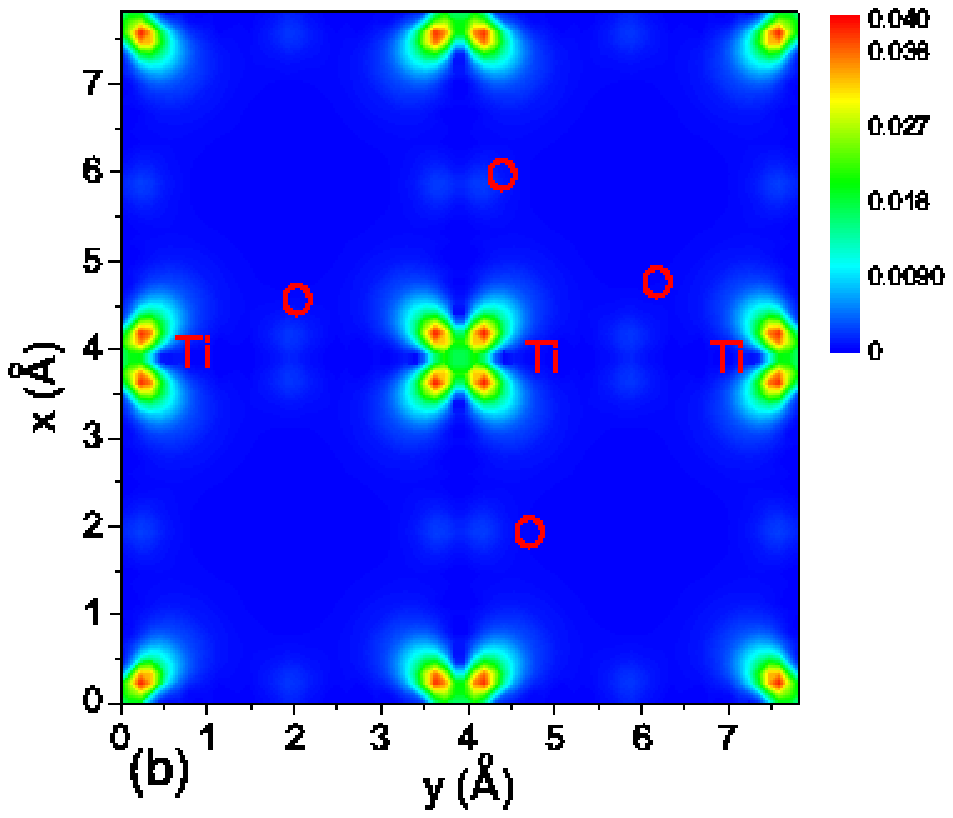}}
\caption{Charge density plots for (a) the AlO$_2$ surface and (b) interface TiO$_2$ layers
with the O-vacancy in the AlO$_2$ surface. Here a supercell LAO(4uc)/STO(4uc) has been considered.
}
\label{fig16}
\end{figure}

Fig.~\ref{fig15} shows the projected densities of states for the surface Al and O atoms and for
the interface Ti. At the AlO$_2$ surface, the local states of a mixed Al$3sp$-O$2p$ character form
in the window ($E_F-1$~eV; $E_F+1$~eV), with the Fermi level
located below the top of the local band.
This implies that the vacancy-doped electrons are mobile. In the interface TiO$_2$ layer,
the vacancy-doped charge occupies the bottom part of the Ti $3d_{xy}$ conducting band, with the distinct
ferromagnetic half-metallic character reported in Ref.~\onlinecite{pavlenko}. A spin polarization 
is also observed
in the AlO$_2$ surface states, the surface magnetic moments being antiparallel to the magnetic
moments of the interface TiO$_2$.

\begin{figure}[htbp]
\epsfxsize=7.5cm {\epsfclipon\epsffile{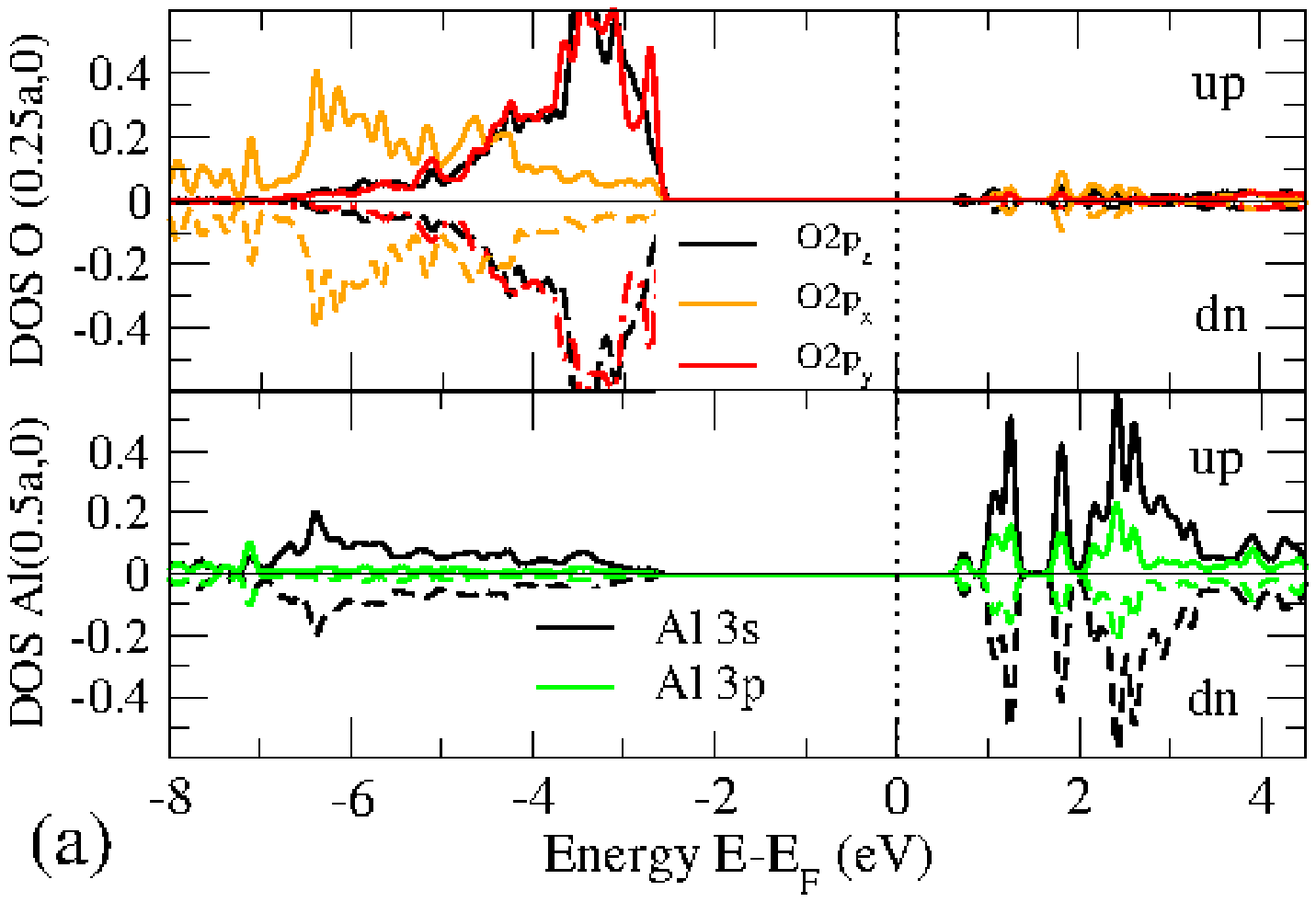}}
\epsfxsize=6.0cm {\epsfclipon\epsffile{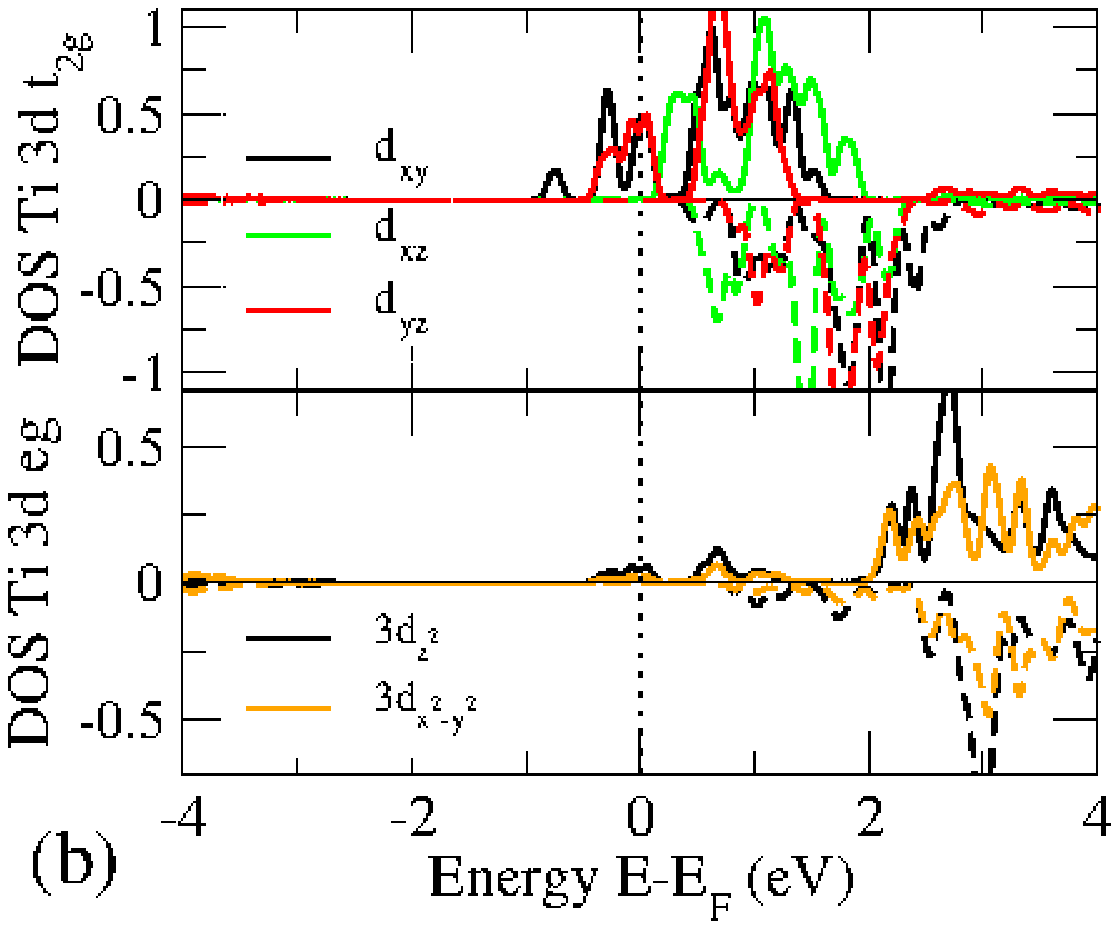}}
\caption{Projected DOS (a) for the surface AlO$_2$ and (b) interface TiO$_2$ layers
in LaAlO$_3$(4u.c.)/SrTiO$_3$(1u.c.) with one O-vacancy per eight O-atoms in the AlO$_2$ surface.
The positions of Ti, Al, La and  O atoms have been fully 
relaxed in the ($x$, $y$)-planes and in the $z$-direction.
}
\label{fig17}
\end{figure}

\begin{figure}[htbp]
\epsfxsize=8.5cm {\epsfclipon\epsffile{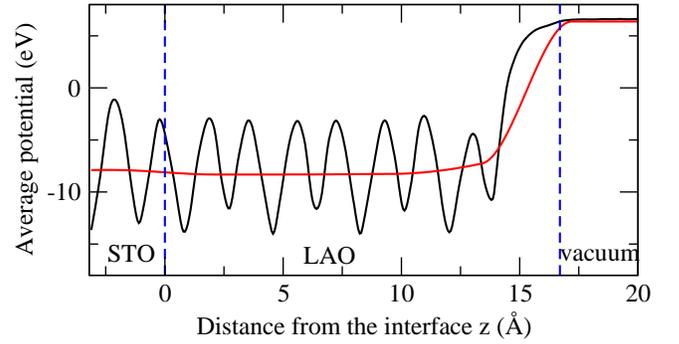}}
\caption{Profile of $xy$-averaged electrostatic potential along the [001] direction 
with $c_V=1/8$-concentration of {\it surface oxygen vacancies}.
The black and red profiles correspond to the $xy$-averaged and macroscopically 
averaged potentials, respectively.}
\label{fig18}
\end{figure}

Fig.~\ref{fig16} shows the spatial distribution of
the conducting charge, both for the charge which arises 
from the Al $3sp$ states and for the charge
in Ti $3d_{xy}$ states. The spatial distribution of the mobile charges at the 
AlO$_2$ surface is presented in Fig.~\ref{fig16}(a). The 
conducting charge in the Al $3sp$-orbitals and O $2p$ orbitals is shifted 
towards the oxygen vacancies which form stripes along the $y$-direction (for the configuration 
of Fig.~\ref{fig1}(b)). In the interface 
TiO$_2$ layer, the conducting electrons reside mostly in the $3d_{xy}$ orbitals 
of Ti atoms (Fig.~\ref{fig16}(b)) which are hybridized with $2p$ states of oxygen atoms. 

For a surface vacancy concentration of $c_V=1/8$, the compensation of the interface polarity
by the vacancy-released electrons leads to a transition of the AlO$_2$ surface to the insulating
state as can be seen in Fig.~\ref{fig17}. The insulating surface state occurs due to the transfer of the
total vacancy-released charge to the interface TiO$_2$ layer.
In the interface TiO$_2$, the transferred charges occupy primarily the $3d$ states  
of Ti atoms nearest to the vacancy. 
The vacancy-generated electron states at the interface are highly spin polarized 
with Ti magnetic moment $M_{Ti}=0.56$~$\mu_B$, enhanced due to a mixed
$e_g$-$t_{2g}$ character of spin polarization indicated in the 
orbital-projected DOS in Fig.~\ref{fig17}(b). 
The electron charge profile for this case is shown in
Fig.~\ref{fig8}(b), which demonstrates the interface 
character of the electrons. 
They are confined mostly to the TiO$_2$ layer near the LAO/STO interface. 
The integration of the charge profile in Fig.~\ref{fig8}(b) reveals a total of two electrons
per oxygen vacancy, which implies a complete suppression of the interface 
polarity by the surface oxygen vacancies, a fact supported also by recent calculations reported 
in Ref.~\onlinecite{yunli}. Fig.~\ref{fig18} demonstrates the calculated $xy$-averaged 
electrostatic potential along the [001]-direction.
In contrast to the biased macroscopic potential in stoichiometric structure (Fig.~\ref{fig9}), 
the macroscopically averaged potential in the structure with the surface vacancies is flat 
in the LAO-layer, which indeed supports the complete suppression of the LAO-polarity 
due to the surface oxygen vacancies.

\section{Conclusions}

Within DFT-calculations, we considered oxygen vacancies in LAO/STO heterostructures and 
performed studies of the orbital states at the LAO/STO interface allowing for several types
of vacancy arrangements. Using generalized gradient approximation (LSDA) with intra-atomic
Coulomb repulsion (GGA+U), we have shown that the oxygen vacancies at the titanate interfaces 
produce a complex multiorbital reconstruction which involves a change of the occupancy of the $e_g$ states
rather than of $t_{2g}$ orbitals of the Ti atoms nearest to the oxygen vacancies.

The orbital reconstruction is accompanied by a magnetic splitting of the local $e_g$
and interface $d_{xy}$ orbitals. This reconstruction generates a two-dimensional magnetic
state not observed in bulk SrTiO$_3$. Moreover, oxygen vacancies placed in the TiO$_2$ 
layer farther away from the interface induce a sizable magnetic moment only in the TiO$_2$ 
interface layer. Also, oxygen vacancies in the AlO$_2$ surface where they are expected 
to be in the most stable configuration, generate a magnetic moment only in the interface 
titanate layer. In this latter case, the electronic reconstruction mechanism due 
to polar catastrophe is suppressed for a vacancy concentration of $c_V\ge 1/8$ by 
the charge introduced by vacancies. The surface is then free of charge carriers 
which implies the formation of an insulating state at the AlO$_2$ surface.

In configurations with vacancy stripes, we have
found an orbital separation of the charge and spin degrees of freedom with 
the vacancy-released charge carriers localized in $e_g$-orbitals, and the spin 
polarization occurring predominantly in $t_{2g}$-states. Moreover, we have 
provided evidence for the generic character
of the two-dimensional magnetic state at titanate surfaces and interfaces.

\section*{Acknowledgements}
This work was  supported by the DFG (TRR~80), Nebraska NSF-EPSCoR (EPS-1010674), and the A.~von~Humboldt Foundation. 
Financial support from the CFI, NSERC, CRC, and from the Max Planck-UBC Centre for 
Quantum Materials is gratefully acknowledged, as are grants of computer time from the UNL Holland Computing 
Center and Leibniz-Rechenzentrum M\"unchen through the SuperMUC project pr58pi.
The authors acknowledge helpful discussion with Rortraud Merkle.

\end{document}